\documentclass[11pt]{article}

\usepackage{amsmath}
\usepackage{amssymb}
\usepackage{latexsym}
\usepackage{epsfig}
\usepackage{hyperref}
\usepackage{color}
\usepackage{amsthm}

\newtheorem*{theorem}{Theorem}
\newtheorem*{corollary}{Corollary}

\def\Tr{\mathop{\rm Tr}\nolimits}

\newcommand{\be}{\begin{eqnarray}}
\newcommand{\ee}{\end{eqnarray}}
\newcommand{\ba}{\begin{array}}
\newcommand{\ea}{\end{array}}
\newcommand{\bq}{\begin{equation}}
\newcommand{\eq}{\end{equation}}
\newcommand{\ft}[2]{{\frac{#1}{#2}}}
\newcommand{\ftb}[2]{{\displaystyle\frac{#1}{#2}}}

\numberwithin{equation}{section}

\begin{document}

\begin{titlepage}
\rightline{August 2010}

\rightline{\tt MIT-CTP-4170}
\begin{center}
\vskip 2.5cm
{\Large \bf {
Invariances and Equations of Motion \\ \vskip 0.25cm in Double Field Theory}}\\
\vskip 2.0cm
{\large {Seung Ki Kwak}}
\vskip 0.5cm
{\it {Center for Theoretical Physics}}\\
{\it {Massachusetts Institute of Technology}}\\
{\it {Cambridge, MA 02139, USA}}\\
sk$\_$kwak@mit.edu

\vskip 2.5cm
{\bf Abstract}

\end{center}

\vskip 0.5cm

\noindent
\begin{narrower}

\noindent We investigate the full set of equations of motion in double field theory and discuss their $O(D,D)$ symmetry and gauge transformation properties. We obtain a Ricci-like tensor, its associated Bianchi identities, and relate our results to those with a generalized metric formulation.

\end{narrower}

\end{titlepage}

\newpage

\tableofcontents
\baselineskip=16pt


\section{Introduction and Summary}
T-duality~\cite{Giveon:1994fu} is a property of string theory that arises from the existence of momentum and winding modes in toroidal compactification. This symmetry is captured by double field theory~\cite{Hull:2009mi}, which doubles the set of coordinates: the usual coordinates $x^i$ associated with momentum states are supplemented by coordinates $\tilde{x}_i$ associated with winding states. The theory describes the massless subsector of closed string theory focusing on gravity $g_{ij}$, antisymmetric tensor $b_{ij}$, and dilaton fields $d$, all of which depend on $x^i$ and $\tilde{x_i}$.
With $i, j = 1, 2, \cdots, D$ double field theory has a continuous $O(D, D)$ symmetry.\footnote{To be more exact, $O(D,D)$ symmetry is the symmetry of double field theory when formulated in non-compact spacetime $\mathbb{R}^{2D}$. For spacetime $\mathbb{R}^{n-1, 1} \times T^d$, where $\mathbb{R}^{n-1, 1}$ is $n$-dimensional Minkowski space and $T^d$ is a torus, the $O(D,D)$ symmetry breaks to $O(n-1,1) \times O(d,d; \mathbb{Z})$. Since our notation covers both the non-compactified and compactified cases, we will refer to the symmetry of either case as $O(D,D)$.}
The constraint $L_0 - \bar{L}_0 =0$ of closed string theory has a field theory implication: $\partial_i \tilde{\partial}^i = 0$ for all gauge parameters and fields. The strong version of this constraint requires that $\partial_i \tilde{\partial}^i = 0$ for all possible products of gauge parameters and fields. Such constraint was imposed in~\cite{Hull:2009zb} and~\cite{Hohm:2010jy}. \\
\\
The background-independent double field theory action~\cite{Hohm:2010jy} is written in terms of ${\cal E}_{ij} = g_{ij} + b_{ij}$ and the dilaton $d$. The action is given by
\bq \label{action}
\ba {c}
S = \displaystyle\int dx d \tilde{x}  \ e^{-2d} \Bigg[ - \ftb{1}{4} g^{ik} g^{jl} {\cal D}^p {\cal E}_{kl} {\cal D}_p {\cal E}_{ij} + \ftb{1}{4} g^{kl} ({\cal D}^j {\cal E}_{ik} {\cal D}^i {\cal E}_{jl} + \bar{{\cal D}}^j {\cal E}_{ki} \bar{\cal D}^i {\cal E}_{lj}) \\ +  ({\cal D}^i d \  \bar{\cal D}^j {\cal E}_{ij} + \bar{\cal D}^i d \ {\cal D}^j {\cal E}_{ji} ) + 4 {\cal D}^i d \ {\cal D}_i d \ \Bigg] ,
\ea
\eq
where the calligraphic derivatives are defined as follows:
\bq \label{calderdef}
{\cal D}_i \equiv \partial_i - {\cal E}_{ik} \tilde{\partial}^k \ , \quad \bar{\cal D}_i \equiv \partial_i + {\cal E}_{ki} \tilde{\partial}^k .
\eq
All upper indices are lowered by $g_{ij} = \ft{1}{2} ({\cal E}_{ij} +{\cal E}_{ji} )$ and lower indices raised by $g^{ij} = (g_{ij})^{-1}$. The action is invariant under (large) $O(D,D)$ transformations that act as follows:
\bq
{\cal E}' (X') =  (a {\cal E}(X) + b) (c {\cal E}(X) + d)^{-1}  \ , \ d' (X') = d(X) \ , \ X' = h X \ , \
h = \left( \begin{array}{cc} a & b \\ c & d  \end{array} \right)  \in O(D,D) .
\eq
The gauge transformations which leave the action invariant are given by
\bq \label{firstgaugetransformation}
\ba{lcl}
\delta {\cal E}_{ij} &=& {\cal D}_i \tilde{\xi}_j - \bar{\cal D}_j \tilde{\xi}_i + \xi^M \partial_M {\cal E}_{ij} + {\cal D}_i {\xi}^k {\cal E}_{kj} + \bar{\cal D}_j \xi^k {\cal E}_{ik} \, , \\ [1.0ex]
\delta d &=& -\ftb{1}{2} \partial_M \xi^M + \xi^M \partial_M d \, ,
\ea
\eq
where ${\xi^M \partial_M = \xi^i \partial_i + \tilde{\xi}_i \tilde{\partial}^i}$ and $\partial_M \xi^M = \partial_i \xi^i + \tilde{\partial}^i \tilde{\xi}_i$\,. \\
Using the strong version of the constraint, the action \eqref{action} is also invariant under the following $\mathbb{Z}_2$ transformation:
\bq \label{z2trf}
{\cal E}_{ij} \rightarrow {\cal E}_{ji} \quad , \quad {\cal D}_i \rightarrow \bar{\cal D}_i \quad , \quad \bar{\cal D}_i \rightarrow {\cal D}_i \quad , \quad d \rightarrow d \, .
\eq
The action \eqref{action} can be rewritten in a manifestly $\mathbb{Z}_2$ symmetric way. (See \eqref{symaction}.)  \\
\\
In this paper we investigate the invariance and the equations of motion of the action \eqref{action}. We focus on the $O(D,D)$ symmetry and the gauge transformations. By varying the action with respect to ${\cal E}$, we first derive the Ricci-like tensor ${\cal K}_{pq}$, whose vanishing is the equation of motion of $\cal E$. The $\cal E$-variation of the manifestly $\mathbb{Z}_2$-symmetrized action gives the $\mathbb{Z}_2$-symmetric ${\cal K}_{pq}$, which is given by
\bq \label{finalkpq}
\ba{lcl}
{\cal K}_{pq} &=& \ftb{1}{4}  (\nabla^s {\cal D}_s {\cal E}_{pq}  + \bar{\nabla}^s \bar{\cal D}_s {\cal E}_{pq} - 2 \nabla ^s {\cal D}_p {\cal E}_{sq} - 2 \bar{\nabla}^s \bar{\cal D}_q {\cal E}_{ps})  -  (\bar{\nabla}_q {\cal D}_p d   + \nabla _p \bar{\cal D}_q d) \\ [2.0ex] 
& &+ \ \ftb{1}{4} g^{jk} (\bar{\cal D}_q {\cal E}_{pj} {\cal D}^i {\cal E}_{ik} + \ftb{1}{2} \bar{\cal D}_q {\cal E}_{ik} {\cal D}^i {\cal E}_{pj} + {\cal D}_p {\cal E}_{jq} \bar{\cal D}^i {\cal E}_{ki}  + \ftb{1}{2} {\cal D}_p {\cal E}_{ki} \bar{\cal D}^i {\cal E}_{jq})  \\ [2.0ex]
& &- \ \ftb{1}{4}  (\bar{\cal D}^j {\cal E}_{iq} {\cal D}^i {\cal E}_{pj} + \ftb{1}{2} \bar{\cal D}^i {\cal E}_{ji} {\cal D}^j {\cal E}_{pq} + \ftb{1}{2} \bar{\cal D}^i {\cal E}_{pq} {\cal D}^j {\cal E}_{ji})  - \ftb{1}{4} g^{ik} g^{jl} \bar{\cal D}_q {\cal E}_{ij} {\cal D}_p {\cal E}_{kl} .
\ea
\eq
Here the $O(D,D)$ covariant derivatives $\nabla$ and $\bar{\nabla}$ are defined as follows~\cite{Hohm:2010jy}: \footnote{These covariant derivatives should be employed when more than one calligraphic derivatives are needed.}
\bq
\ba{lcl}
\nabla_i \bar{\eta}_j & \equiv & {\cal D}_i \bar{\eta}_j - \Gamma_{i \bar{j}}^{\bar{k}} \, \bar{\eta}_k \, , \\ [1.0ex]
\bar{\nabla}_j {\eta}_i & \equiv & \bar{\cal D}_j {\eta}_i - \Gamma_{ \bar{j} i }^{k} \, {\eta}_k \, ,
\ea
\eq
where the Christoffel-like symbols are defined by
\bq
\ba{lclcl}
\Gamma_{i \bar{j}}^{\bar{k}} &=& \ftb{1}{2} g^{kl} \Big( {\cal D}_i {\cal E}_{lj} + \bar{\cal D}_j {\cal E}_{il} - \bar{\cal D}_l {\cal E}_{ij}   \Big) \ , \quad \Gamma_{ij}^k &=& \ftb{1}{2} g^{kl}  {\cal D}_i {\cal E}_{jl} \, , \\ [2.0ex]
\Gamma_{\bar{i} j}^{k} &=& \ftb{1}{2} g^{kl} \Big( \bar{\cal D}_i {\cal E}_{jl} + {\cal D}_j {\cal E}_{li} - {\cal D}_l {\cal E}_{ji}   \Big) \ , \quad \Gamma_{\bar{i} \bar{j}}^{\bar{k}} &=& \ftb{1}{2} g^{kl}  \bar{\cal D}_i {\cal E}_{lj} \, .
\ea
\eq
The theory has both unbarred and barred indices. The first index of ${\cal E}_{ij}$ is unbarred and the second index is barred. The index of ${\cal D}_i$ is unbarred while the index of $\bar{\cal D}_i$ is barred. The indices of $g^{ij}$ are either both unbarred or both barred. Throughout this paper, for notational simplicity we will not explicitly display whether an index is unbarred or barred. Two different types of indices transform differently under the $O(D,D)$ transformation thus contractions between same-type indices are $O(D,D)$ covariant. It is manifest that ${\cal K}_{pq}$ is $O(D,D)$ covariant from its index structure. However, it is generally not true that the $\cal E$-variation of an $O(D,D)$ tensor is an $O(D,D)$ tensor. Terms obtained from the direct $\cal E$-variation of the action \eqref{action} are not manifestly $O(D,D)$ covariant. They were manipulated to assemble in the manifestly $O(D,D)$ covariant form shown above.\\
\\
We derive the gauge transformation of ${\cal K}_{pq}$ from the gauge invariance of the action. The result is as follows:
 \bq \label{firstfullgt}
 \delta {\cal K}_{pq} = {\cal L}_{\xi} {\cal K}_{pq} + {\cal L}_{\tilde{\xi}} {\cal K}_{pq} + {\cal E}_{pk} (\tilde{\partial}^l \xi^k - \tilde{\partial}^k \xi^l ) {\cal K}_{lq} + {\cal E}_{lq} (\tilde{\partial}^l \xi^k - \tilde{\partial}^k \xi^l ) {\cal K}_{pk} \, .
 \eq
 This result is rewritten in terms of $O(D,D)$ covariant derivatives and parameters in \eqref{bi_gt}. From this we infer that the gauge transformation of ${\cal K}_{pq}$ is manifestly $O(D,D)$ \textit{non-covariant}. \\
 \\
 Contracted Bianchi identities can be derived from the gauge invariance of the action. We find
 \be
\bigg[ {\nabla}^p - \ftb{1}{2} g^{ps} (\bar{\cal D}^r {\cal E}_{sr} + 4 {\cal D}_s d) \bigg] {\cal K}_{pq} + \ftb{1}{2} \bar{\cal D}_q {\cal R} &=& 0 \label{final_bi1} \, , \\
\bigg[ \bar{\nabla}^q - \ftb{1}{2} g^{qs} ({\cal D}^r {\cal E}_{rs} + 4 \bar{\cal D}_s d) \bigg] {\cal K}_{pq} + \ftb{1}{2} {\cal D}_p {\cal R} &=& 0  \label{final_bi2} \, .
\ee
These two Bianchi identities both reduce to the Bianchi identity in General Relativity when there is no $\tilde{x}$ dependence. In this limit the equations of motion ${\cal R}=0$ and ${\cal K}_{pq}=0$ also reduce to the equations of motion in General Relativity with $g_{ij}$, $b_{ij}$ and a scalar field $\phi$, where $e^{-2d} \equiv \sqrt{-g} e^{-2 \phi}$. This setup will be referred to as a conventional GR setup throughout this paper: the equations of motion and the Bianchi identity in this setup can be found in Appendix \ref{usualgr}.\\
\\
Our results are closely related to those in a generalized metric formulation~\cite{Hohm:2010pp}. In this formulation the action \eqref{action} is written in terms of the generalized metric ${\cal H}^{MN}$ instead of ${\cal E}_{pq}$. The direct variation of the action with respect to ${\cal H}^{MN}$ gives the tensor ${\cal K}_{MN}$, whose vanishing is not the equation of motion since $\cal H$ is a constrained field. ${\cal R}_{MN}$ is defined from ${\cal K}_{MN}$ by implementing the constrained variation, thus the vanishing of ${\cal R}_{MN}$ gives the correct equation of motion. The gauge transformations of ${\cal R}_{MN}$ or ${\cal K}_{MN}$ were not explicitly discussed in~\cite{Hohm:2010pp}. We first prove that ${\cal R}_{MN}$ is a generalized tensor while ${\cal K}_{MN}$ is not. Following~\cite{Siegel:1993th}, reference~\cite{Hohm:2010pp} also discussed the vielbein formulation of the double field theory whose frame fields $e^M{}_A$ implement a local $GL (D, \mathbb{R})
  \times GL (D, \mathbb{R})$ symmetry. By gauge-fixing this symmetry, this formulation can be written in terms of $\cal E$. We show that ${\cal R}_{a \bar{b}} \equiv e^M{}_a {\cal R}_{MN} e^N{}_{\bar{b}}$ exactly coincides with $- {\cal K}_{a \bar{b}} $ given by \eqref{finalkpq} after appropriate gauge fixing of the local $GL (D, \mathbb{R}) \times GL (D, \mathbb{R})$ symmetry. Finally using the gauge transformation property of a generalized tensor, which is given by the generalized Lie derivatives $\hat{\cal L}_{\xi}$, we find another way to derive the gauge transformation \eqref{firstfullgt} of ${\cal K}_{pq}$. \\
\\
For many derivations in this paper, the invariances of the theory play crucial roles. The $O(D,D)$ invariance of the action directly gives the result that ${\cal K}_{pq}$ is $O(D,D)$ covariant. The gauge transformation of ${\cal K}_{pq}$ can be derived from the gauge invariance of the action. Even in the generalized metric formulation, gauge invariance is a key to understand that ${\cal R}_{MN}$ is a generalized tensor.
\section{$\cal E$-variation of Double Field Theory Action}
In this section we start with the review of early works to provide definitions and formulas that we need for the rest of this paper. We then vary the double field theory action with respect to $\cal E$ to obtain ${\cal K}_{pq}$, whose explicit form was given in \eqref{finalkpq}.
${\cal K}^{pq}$ is defined by varying the action with respect to ${\cal E}_{pq}$ as follows:
\bq \label{defk}
\delta_{{\cal E}} S = \displaystyle\int dx d \tilde{x}  \ e^{-2d} \delta {\cal E}_{pq} {\cal K}^{pq} \ .
\eq
Then ${\cal K}_{pq}$ is defined by lowering indices of ${\cal K}^{pq}$ with $g_{ij}$, i.e.  ${\cal K}_{pq} \equiv g_{pk} g_{ql} {\cal K}^{kl}$.
The $O(D,D)$ covariance of this tensor ${\cal K}_{pq}$ will be shown from the $O(D,D)$ invariance of the action and from the theorem about the general $O(D,D)$ structure of the $\cal E$-variation of an $O(D,D)$ tensor. We conclude this section by displaying the relation of ${\cal K}_{pq}$ to the Ricci tensor.
\subsection{Review of previous works} \label{review}
Following the early work of Tseytlin~\cite{Tseytlin:1990nb} and Siegel~\cite{Siegel:1993th, Siegel:1993xq}, the work in~\cite{Hull:2009mi} constructed the perturbative double field theory action up to cubic order. By imposing the strong version of the constraint,  it was pointed out in~\cite{Hull:2009zb} that the gauge algebra of double field theory reduces to the Courant bracket and more recent work in~\cite{Hohm:2010jy} provided the fully background-independent double field theory action. The most recent paper~\cite{Hohm:2010pp} introduced the generalized metric formulation and found much simpler expressions for the double field theory action and the Ricci-like tensor. Some definitions and formulas needed for the rest of this paper will be displayed below.\\
\\
To achieve background independence, the strong version of the constraint was imposed~\cite{Hohm:2010jy}: $\partial^M \partial_M$ kills all possible products of fields and gauge parameters in the theory i.e. $\partial^M \partial_M (AB) =0$ for any $A$ and $B$.\footnote{By imposing this constraint, the field space is truncated into the null subspace of double field theory and the gauge symmetry reduces to diffeomorphism and $b$-field gauge transformations on that subspace.} The constraint is equivalent to $\partial^M A \partial_M B =0$ and this can be written in terms of calligraphic derivatives as follows:
\bq \label{strongcon}
{\cal D}_i A {\cal D}^i B = \bar{\cal D}_i A \bar{\cal D}^i B.
\eq
We mentioned that there are two different types of indices transforming differently under the $O(D,D)$ transformation. More generally an $O(D,D)$ tensor with lower indices $T_{i_1 \cdots i_p , \bar{j}_1 \cdots \bar{j}_1}$ transforms as follows:
\bq \label{lowertensor}
T_{i_1 \cdots i_p , \bar{j}_1 \cdots \bar{j}_q} (X) = M_{i_1}{}^{k_1} \cdots M_{i_p}{}^{k_p} \bar{M}_{\bar{j}_1}{}^{\bar{l}_1} \cdots \bar{M}_{\bar{j}_q}{}^{\bar{l}_q}
T'_{k_1 \cdots k_p , \bar{l}_1 \cdots \bar{l}_q} (X') \, ,
\eq
and an $O(D,D)$ tensor with upper indices $U^{i_1 \cdots i_p , \bar{j}_1 \cdots \bar{j}_q}$ transforms as
\bq \label{uppertensor}
U^{i_1 \cdots i_p , \bar{j}_1 \cdots \bar{j}_q} (X) = U'^{k_1 \cdots k_p , \bar{l}_1 \cdots \bar{l}_q} (X') \ (M^{-1})_{k_1}{}^{i_1} \cdots (M^{-1})_{k_p}{}^{i_p} (\bar{M}^{-1})_{\bar{l}_1}{}^{\bar{j}_1} \cdots (\bar{M}^{-1})_{\bar{l}_q}{}^{\bar{j}_q}.
\eq
Here $M$ and $\bar{M}$ are defined by
\bq
M(X) \equiv d^t - {\cal E} (X) c^t \  , \quad \bar{M}(X) \equiv d^t + {\cal E}^t (X) c^t \, .
\eq
While the matrices $M$ and $M^{-1}$ depend on $X$ (not $X'$), this dependence is omitted in \eqref{lowertensor} and \eqref{uppertensor}. We will not display the $X$ dependence of $M$ and $M^{-1}$ explicitly throughout the paper. \\
\\
In~\cite{Hohm:2010pp} the double field theory action \eqref{action} is rewritten in terms of the generalized metric ${\cal H}$ and the dilaton $d$. The action in this formulation is given by
\bq \label{genmetaction}
\ba{c}
S =  \displaystyle\int dx d \tilde{x}  \ e^{-2d} \Bigg( \ftb{1}{8} {\cal H}^{MN} \partial_M {\cal H}^{KL} \partial_N {\cal H}_{KL} - \ftb{1}{2}  {\cal H}^{MN} \partial_N {\cal H}^{KL} \partial_L {\cal H}_{MK} \\
-2 \partial_M d \partial_N {\cal H}^{MN} + 4  {\cal H}^{MN} \partial_M d \partial_N d \Bigg) \, .
\ea
\eq
The explicit form of ${\cal H}$ is written as
\bq
{\cal H}_{MN} = \left( \begin{array}{cc} g^{ij} & -g^{ik}b_{kj} \\ [1.0ex] b_{ik} g^{kj} & g_{ij} - b_{ik} g^{kl} b_{lj} \end{array} \right)  \, .
\eq
${\cal K}_{MN}$ is defined by varying the action with respect to ${\cal H}$:
\bq \label{hvariation}
\delta_{\cal H} S =  \displaystyle\int dx d \tilde{x}  \ e^{-2d} \delta {\cal H}^{MN} {\cal K}_{MN} \, .
\eq
The explicit form of ${\cal K}_{MN}$ is given by~\cite{Hohm:2010pp}
\bq \label{kmn}
\ba{lcl} 
{\cal K}_{MN} &=& \ftb{1}{8} \partial_M {\cal H}^{KL} \partial_N {\cal H}_{KL} - \ftb{1}{4} (\partial_L - 2(\partial_L d)) ({\cal H}^{LK} \partial_K {\cal H}_{MN}) + 2 \partial_M \partial_N d \\ [2.0ex]
&&- \ftb{1}{2}  \partial_{(M} {\cal H}^{KL} \partial_L {\cal H}_{N)K} + \ftb{1}{2} (\partial_L - 2(\partial_L d)) ({\cal H}^{KL} \partial_{(M} {\cal H}_{N)K} + {{\cal H}^K}_{(M} \partial_K {{\cal H}^L}_{N)} ) \, ,
\ea
\eq
where indices are lowered and raised with the constant $O(D,D)$ invariant metric $\eta_{MN} = \begin{pmatrix} 0&  1 \\1 & 0\end{pmatrix} $. \\
\\
$\cal H$ is a constrained field satisfying ${\cal H} \eta {\cal H} = \eta^{-1}$. By implementing the constraint for ${\cal H}$ in the variation \eqref{hvariation}, ${\cal R}_{MN}$ is defined by (4.57) of~\cite{Hohm:2010pp} (in matrix form):
\bq \label{ma_form}
{\cal R} \equiv  \ftb{1}{4} (1- S^t) {\cal K} (1+S) + \ftb{1}{4} (1+ S^t) {\cal K} (1-S) \, .
\eq
A generalized tensor $A_M{}^N$ is defined in~\cite{Hohm:2010pp} to have the following gauge transformation properties:
\bq
\delta_{\xi} A_M{}^N = \hat{\cal L}_{\xi} A_M{}^N \, ,
\eq
where the generalized Lie derivatives $\hat{\cal L}$ are defined as follows:
\bq \label{genlied} 
\hat{\cal L}_{\xi} A_M{}^N \equiv \xi^P \partial_P A_M{}^N + (\partial_M \xi^P - \partial^P \xi_M ) A_P{}^N + ( \partial^N \xi_P  - \partial_P \xi^N) A_M{}^P.
\eq
The definitions and formulas in the vielbein formulation will be introduced in section \ref{genmetsec} since they are only needed in the last parts of the section.
\subsection{$O(D,D)$ covariance of ${\cal K}_{pq}$} \label{argueodd}
T-duality is the well-known global symmetry of closed string theory on a torus.  This symmetry becomes enlarged to a continuous $O(D,D)$ symmetry for the massless subsector that we are considering. The $O(D,D)$ symmetry is implemented in the double field theory action \eqref{action}. Here we demonstrate the $O(D, D)$ covariance of ${\cal K}_{pq}$ using the $O(D,D)$ invariance of the double field theory action. \\
\\
For some given background field ${\cal E}$, one can define an arbitrary variation $\delta {\cal E}$ such that ${\cal E}' = {\cal E} + \delta {\cal E}$. Consider an $O(D,D)$ transformation $h$ as follows:
\bq
\ba{lcl}
h = \left( \begin{array}{cc} a & b \\ c & d  \end{array} \right) &:& \begin{array}{l} {\cal E} \rightarrow {\cal E}_h = (a {\cal E} + b) (c {\cal E} + d)^{-1} \ , \quad {\cal E}' \rightarrow {\cal E}'_h = (a {\cal E}' + b) (c {\cal E}' + d)^{-1} \, ,  \\ [0.5ex]
  X \rightarrow X_h = h  X  \ , \quad d \rightarrow d_h = d  \, . \end{array} 
\ea
\eq
While ${\cal E}_{pq}$ is not an $O(D,D)$ covariant object, the variation $\delta {\cal E}_{pq}$ is.~\cite{Hohm:2010jy} We can thus use \eqref{lowertensor} for the $O(D,D)$ transformation of $\delta {\cal E}$ as follows:
\bq
\delta {\cal E}_h (X_h ) \equiv {\cal E}'_h (X_h) - {\cal E}_h (X_h) = M^{-1} \delta {\cal E} (X) ( \bar{M}^t )^{-1}  \, .
\eq
The definition of ${\cal K}^{pq}$ \eqref{defk} gives
\bq \label{diffaction1}
S [{\cal E}', d] - S [{\cal E}, d]  = \displaystyle\int dx d \tilde{x}  \ e^{-2d} \delta {\cal E}_{pq} {\cal K}^{pq} = \displaystyle\int dx d \tilde{x}  \ e^{-2d} \Tr \Big[ \big( \delta {\cal E}  (X) \big)^t {\cal K} (X) \Big] \, .
\eq
Under the $O(D,D)$ transformation $h$, ${\cal K}$ transforms to ${\cal K}_h$. Then the following equality holds for the transformed action:
\bq \label{diffaction2}
\ba{lcl}
S [{\cal E}'_h, d] - S [{\cal E}_h, d]  &=&  \displaystyle\int dx_h d \tilde{x}_h  \ e^{-2d}  \Tr \Big[ \big( \delta {\cal E}_h  (X_h ) \big)^t {\cal K}_h (X_h ) \Big]  \\ [2.0ex]
&=& \displaystyle\int dx_h d \tilde{x}_h  \ e^{-2d} \Tr \Big[ \bar{M}^{-1}  \big(  \delta {\cal E} (X) \big)^t  (M^t)^{-1}  {\cal K}_h (X_h) \Big] \, .
\ea
\eq
The $O(D,D)$ invariance implies that the left hand side of \eqref{diffaction1} and \eqref{diffaction2} are equal. Since $dx d \tilde{x} = \det [ h ] dx_h d \tilde{x}_h = dx_h d \tilde{x}_h$ (measure is invariant under the $O(D,D)$ transformation), this equality gives
\bq
\ba{lcl}
0 &=&  \displaystyle\int dx d \tilde{x}  \ e^{-2d} \Tr \Big[ \big( \delta {\cal E}  (X) \big)^t {\cal K} (X) - \bar{M}^{-1}  \big(  \delta {\cal E} (X) \big)^t  (M^t)^{-1}   {\cal K}_h (X_h) \Big] \\ [2.0ex]
&=&  \displaystyle\int dx d \tilde{x}  \ e^{-2d}  \Tr \Big[ \big( \delta {\cal E}  (X) \big)^t \Big\{ {\cal K} (X) -  (M^t)^{-1}   {\cal K}_h (X_h)  \bar{M}^{-1} \Big\} \Big] 
\ea
\eq
Since $\delta {\cal E}$ is arbitrary, this shows the $O(D,D)$ covariance of ${\cal K}$:
\bq
{\cal K} (X) =  (M^t)^{-1}   {\cal K}_h (X_h)  \bar{M}^{-1}  \, .
\eq
The first index of ${\cal K}^{pq}$ transforms with $M^{-1}$ and the second index transforms with $\bar{M}^{-1}$. With \eqref{uppertensor} we conclude that ${\cal K}^{pq}$ is an $O(D,D)$ tensor with two upper indices, where the first index is unbarred while the second index is barred. From the definition, ${\cal K}_{pq}$ is obtained by lowering indices of ${\cal K}^{pq}$ with $g_{ij}$, hence ${\cal K}_{pq}$ is also an $O(D,D)$ covariant tensor, with the first index unbarred and the second index barred.
\subsection{Alternative explanation of $O(D,D)$ covariance} \label{altodd}
The $\cal E$-variation of an $O(D,D)$ tensor is generally not an $O(D, D)$ tensor. This can be seen from the following simple example:
\bq
\delta_{\cal E} {\cal D}_i d = - \delta {\cal E}_{ik} \tilde{\partial}^k d = \ftb{1}{2} \delta {\cal E}_{ik} (\underline{{\cal D}^k d} - \bar{\cal D}^k d ) \, ,
\eq
where the underlined term is $O(D,D)$ non-covariant because of the inappropriate contraction between a barred and an unbarred index.
However, an object resulting from the $\cal E$-variation of  an $O(D,D)$ tensor has some general $O(D,D)$ structure. In this section a theorem on the general $O(D,D)$ structure of $\cal E$-variations will be proven. The $O(D,D)$ covariance of ${\cal K}_{pq}$ follows from this theorem. A detailed proof of the theorem can be found in Appendix \ref{genoddthm1}.
\begin{theorem}
The ${\cal E}$-variation of any $O(D,D)$ tensor $T$ (with both upper and lower indices) has the following $O(D,D)$ structure:
\bq \label{tensor_realprop}
\ba{lcl}
\delta_{\cal E} T_{i_1 i_2 \cdots i_n}^{i_1' i_2' \cdots i_{n'}'} &=& \!\!\!\!\!\!  \displaystyle\sum_{\substack{\text{all unbarred} \\ \text{lower indices $k$}}} \!\! \ftb{1}{2} \delta {\cal E}_{kq} \, g^{ql} \, T_{\{ k \rightarrow l \}} + \!\!\!\!\!\!  \displaystyle\sum_{\substack{\text{all barred} \\ \text{lower indices $k$}}} \!\! \ftb{1}{2} \delta {\cal E}_{pk} \, g^{pl} \, T_{\{ k \rightarrow l\}} \\
 &&- \!\!\!\!\!\! \displaystyle\sum_{\substack{\text{all unbarred} \\ \text{upper indices $k$}}} \!\! \ftb{1}{2} \delta {\cal E}_{pq} \, g^{qk} \, T^{\{ k \rightarrow p \}}
- \!\!\!\!\!\! \displaystyle\sum_{\substack{\text{all barred} \\ \text{upper indices $k$}}} \!\! \ftb{1}{2} \delta {\cal E}_{pq} \, g^{pk} \, T^{\{ k \rightarrow q\}} + \Big( O(D,D) \Big) \, ,
\ea
\eq
where $T_{ \{ k \rightarrow l \} }$ is the tensor $T$ with the lower index $k$ replaced by the index $l$ and $T^{ \{ k \rightarrow l \} }$ is the tensor $T$ with the upper index $k$ replaced by the index $l$. The sums above are $O(D,D)$ non-covariant terms. The term $\big( O(D,D) \big)$ represents $O(D,D)$ covariant terms. \footnote{One feature of this theorem is that non-covariant terms are invariant under the $\mathbb{Z}_2$ transformation \eqref{z2trf} if the tensor $T$ is invariant under $\mathbb{Z}_2$ transformation. In case $T$ does not have the $\mathbb{Z}_2$ symmetry, the structure of non-covariant terms is still preserved.}
\end{theorem}
\noindent \textit{Proof}. See Appendix \ref{genoddthm1}. \\
\\
This theorem states that the ${\cal E}$-variation of an $O(D,D)$ covariant tensor generally has $O(D,D)$ non-covariant terms. Since there is no index for an $O(D,D)$ scalar, by the theorem, the ${\cal E}$-variation of an $O(D,D)$ scalar does not have any $O(D,D)$ non-covariant term. This leads to the following corollary:

\begin{corollary}
The ${\cal E}$-variation of any $O(D,D)$ scalar is an $O(D, D)$ scalar.
\end{corollary}
\noindent The Lagrangian $\mathcal{L}$ of the double field theory is an $O(D,D)$ scalar thus the $\cal E$-variation of the Lagrangian must be an 
$O(D,D)$ scalar.  
The direct ${\cal E}$-variation of the Lagrangian $\mathcal{L}$ takes the following form:
\bq \label{evarlag}
\delta_{\cal E} {\mathcal{L}} = \delta {\cal E}_{pq} T_0^{pq} + \nabla_i ( \delta {\cal E}_{pq}) T_1^{ipq} +  \bar{\nabla}_i ( \delta {\cal E}_{pq}) T_2^{ipq} \, .
\eq
Since $ \delta {\cal E}_{pq},  \nabla_i ( \delta {\cal E}_{pq})$, and 
$ \bar{\nabla}_i ( \delta {\cal E}_{pq})$ 
are linearly independent and the sum of the terms above is $O(D,D)$ covariant, each $T_k$ for $k=0,1,2$ must be separately $O(D,D)$ covariant. The definition \eqref{defk} of ${\cal K}^{pq}$ requires the integration by parts for the second and the third terms in \eqref{evarlag}. 
The above $\cal E$-variation of the Lagrangian can thus be rewritten as
\bq
\delta_{\cal E} {\mathcal{L}}  = \delta {\cal E}_{pq} \bigg[ T_0^{pq} - \nabla_i T_1^{ipq}  -   \bar{\nabla}_i T_2^{ipq} + \ftb{1}{2}  \bar{\cal D}^k {\cal E}_{ik} T_1^{ipq}   +   \ftb{1}{2} {\cal D}^k {\cal E}_{ki}  T_2^{ipq}  \bigg]  + (\text{total derivatives}) \, ,
\eq
where each term within the brackets is an $O(D,D)$ tensor.
The object in  brackets is, in fact,  ${\cal K}^{pq}$ thus we conclude that 
${\cal K}^{pq}$ is an $O(D,D)$ covariant tensor. The $O(D,D)$ covariance of $g^{ij}$ then leads to the result that ${\cal K}_{pq}$ is an $O(D,D)$ tensor. Furthermore, since each term of the action (either \eqref{action} or the $\mathbb{Z}_2$-symmetrized action \eqref{symaction}) is an $O(D,D)$ scalar, the $\cal E$-variation of each term should separately yield $O(D,D)$ covariant terms. This can be explicitly seen in Appendix \ref{appA}. \\
\\
The origin of $O(D,D)$ non-covariant terms in the $\cal E$-variation of an $O(D,D)$ tensor is from the variation of the inverse metric $g^{ij}$, calligraphic derivatives ${\cal D}$, $\bar{\cal D}$, and covariant derivatives ${\nabla}$, $\bar{\nabla}$. Hence it follows that the $d$-variation of an $O(D,D)$ tensor does not yield such $O(D,D)$ non-covariant terms. These properties of the $\cal E$-variation and the $d$-variation of an $O(D,D)$ tensor will be used in section \ref{gaugetr section} to check the form of the gauge transformation of ${\cal K}_{pq}$.
\subsection{Calculation of ${\cal K}_{pq}$}
The manifestly $\mathbb{Z}_2$ symmetric form of the action \eqref{action} can be written as follows:
\bq \label{symaction}
\ba {lcl}
S &=& \displaystyle\int dx d \tilde{x}  \ e^{-2d} \Bigg[ - \ftb{1}{4} g^{ik} g^{jl}  \Big( \ftb{1}{2}  {\cal D}^p {\cal E}_{kl} {\cal D}_p {\cal E}_{ij} + \ftb{1}{2}  \bar{\cal D}^p {\cal E}_{kl} \bar{\cal D}_p {\cal E}_{ij}  \Big)  + \ftb{1}{4} g^{kl} ({\cal D}^j {\cal E}_{ik} {\cal D}^i {\cal E}_{jl} \\ 
&& + \ \bar{{\cal D}}^j {\cal E}_{ki} \bar{\cal D}^i {\cal E}_{lj})  +  ({\cal D}^i d \  \bar{\cal D}^j {\cal E}_{ij} + \bar{\cal D}^i d \ {\cal D}^j {\cal E}_{ji} ) + 2 ( {\cal D}^i d \ {\cal D}_i d + \bar{\cal D}^i d \ \bar{\cal D}_i d) \ \Bigg] \, .
\ea
\eq
This action is equivalent to \eqref{action} using the constraint \eqref{strongcon}.
 The ${\cal E}$-variation of the action \eqref{symaction} gives:
\bq \label{kpq3}
\ba{lcl}
{\cal K}^{pq} &=&  \ftb{1}{4} g^{ip} g^{jq} (\nabla^s {\cal D}_s {\cal E}_{ij}  + \bar{\nabla}^s \bar{\cal D}_s {\cal E}_{ij} - 2 \nabla ^s {\cal D}_i {\cal E}_{sj} - 2 \bar{\nabla}^s \bar{\cal D}_j {\cal E}_{is})  - g^{pi} g^{qj} (\bar{\nabla}_j {\cal D}_i d   + \nabla _i \bar{\cal D}_j d) \\ [2.0ex]
&&+ \ \ftb{1}{4} g^{pl} g^{jk} (\bar{\cal D}^q {\cal E}_{lj} {\cal D}^i {\cal E}_{ik} + \ftb{1}{2} \bar{\cal D}^q {\cal E}_{ik} {\cal D}^i {\cal E}_{lj})  + \ftb{1}{4} g^{ql} g^{jk} ({\cal D}^p {\cal E}_{jl} \bar{\cal D}^i {\cal E}_{ki}  + \ftb{1}{2} {\cal D}^p {\cal E}_{ki} \bar{\cal D}^i {\cal E}_{jl})  \\ [2.0ex]
&&- \ \ftb{1}{4} g^{pk} g^{ql} (\bar{\cal D}^j {\cal E}_{il} {\cal D}^i {\cal E}_{kj} + \ftb{1}{2} \bar{\cal D}^i {\cal E}_{ji} {\cal D}^j {\cal E}_{kl} + \ftb{1}{2} \bar{\cal D}^i {\cal E}_{kl} {\cal D}^j {\cal E}_{ji}) 
     - \ftb{1}{4} g^{ik} g^{jl} \bar{\cal D}^q {\cal E}_{ij} {\cal D}^p {\cal E}_{kl} \, ,
\ea
\eq
using the integration by parts. This ${\cal K}^{pq}$ is equivalent to that obtained from the $\cal E$-variation of the action \eqref{action}.
To see this more clearly, consider the constraint \eqref{strongcon}.
By taking the ${\cal E}$-variation on both sides, one obtains
\be
\delta_{\cal E} \Big( {\cal D}_i A {\cal D}^i B \Big) = {\cal D}_i \big( \delta_{\cal E} A \big) {\cal D}^i B +  {\cal D}_i A {\cal D}^i \big( \delta_{\cal E}B \big) - \ftb{1}{2} \delta {\cal E}_{pq} \Big[ {\cal D}^p A \bar{\cal D}^q B +  \bar{\cal D}^q A {\cal D}^p B \Big] \, , \\
\delta_{\cal E} \Big( \bar{\cal D}_i A \bar{\cal D}^i B \Big) = \bar{\cal D}_i \big( \delta_{\cal E} A \big) \bar{\cal D}^i B +  \bar{\cal D}_i A \bar{\cal D}^i \big( \delta_{\cal E}B \big) - \ftb{1}{2} \delta {\cal E}_{pq} \Big[ {\cal D}^p A \bar{\cal D}^q B +  \bar{\cal D}^q A {\cal D}^p B \Big] \, ,
\ee
which are equivalent up to the constraint. This justifies the equivalence of ${\cal K}^{pq}$ obtained from the actions \eqref{action} and \eqref{symaction}. \\
\\
Under the $\mathbb{Z}_2$ transformation, ${\cal K}^{pq}$ defined in \eqref{kpq3} transforms as follows:
\bq
{\cal K}^{pq} \rightarrow {\cal K}^{qp} \, , \nonumber
\eq
which implies that the field equation ${\cal K}_{pq}=0$ is $\mathbb{Z}_2$ invariant. This symmetry becomes manifest since we have the following $\mathbb{Z}_2$ transformation properties of $\nabla$ and $\bar{\nabla}$:
\bq
\nabla_i \rightarrow \bar{\nabla}_i \ , \quad \bar{\nabla}_i \rightarrow \nabla_i \, .
\eq
We obtain \eqref{finalkpq} by lowering indices of \eqref{kpq3} using $g_{ij}$, which concludes the determination of ${\cal K}_{pq}$. All detailed calculations are included in Appendix \ref{appA}. \\
\\
From the arguments in section \ref{argueodd} and \ref{altodd}, ${\cal K}^{pq}$ given in \eqref{kpq3} is an $O(D,D)$ covariant tensor. Moreover, it can be easily seen that each term in ${\cal K}^{pq}$ is manifestly $O(D,D)$ covariant. This is due to the contractions between appropriate indices i.e. contractions between barred (unbarred) indices and barred (unbarred) indices respectively. \\
\\
When there is no $\tilde{x}$ dependence, ${\cal K}_{pq}$ should reduce to an object which transforms covariantly under diffeomorphisms in a conventional GR setup. With some lengthy calculation, we obtain the following explicit form in this limit:
\bq
\ba{l} \label{kpqlimit}
{\cal K}_{pq} \Big|_{\tilde{\partial}=0} = \Bigg[ - R_{pq}  + \ftb{1}{4} {H_{p}}^{r s} H_{q r s}  - 2 \nabla_p \nabla_q \phi \Bigg] + \Bigg[ \ftb{1}{2} \nabla^s H_{spq} - H_{spq} \nabla^s \phi \Bigg] \, ,
\ea
\eq
where $R_{pq}$ is the Ricci tensor, $H_{ijk} \equiv 3 \partial_{[i} b_{jk]} $ is the field strength, and $\phi$ is the scalar dilaton. Note that the terms in the first square bracket are symmetric under the exchange of $p$ and $q$, while the terms in the second square bracket are antisymmetric under the same exchange of indices.
\section{Bianchi Identities and Equations of Motion}
A contracted Bianchi identity can be derived in a few different ways. One derivation is starting from the Bianchi identity of the Riemann curvature tensor and then contracting it. Since we have no notion of a Riemann-type tensor in double field theory so far, we will use another way to derive a Bianchi identity, which uses the gauge symmetry of the action. \\
\\
The equations of motion in double field theory are simply ${\cal R}=0$ and ${\cal K}_{pq} =0$: the former is the equation of motion for dilaton $d$ and the latter is the equation of motion for ${\cal E}$. Since ${\cal R}$ was explicitly derived in~\cite{Hohm:2010jy} and the limit case of no $\tilde{x}$ dependence was investigated in the same paper, we will focus on ${\cal K}_{pq}$ especially in the limit of no $\tilde{x}$ dependence.
Our results in the limit of no $\tilde{x}$ dependence will be compared with those in a conventional GR setup, which can be found in Appendix \ref{usualgr}.
\subsection{Derivation of Bianchi identities}
Bianchi identities in double field theory can be derived from the gauge symmetry of the action \eqref{action}. The gauge transformations of ${\cal E}$ and $d$ are given by the following manifestly $O(D,D)$ covariant form:
\be 
\delta {\cal E}_{ij} &=& \nabla_i \bar{\eta}_j + \bar{\nabla}_j \eta_i   \label{gtform1} \, , \\
\delta d &=& - \ftb{1}{4} \bigg( \nabla_i - \ftb{1}{2} (\bar{\cal D}^p {\cal E}_{ip} + 4 {\cal D}_i d ) \bigg) \eta^i 
-  \ftb{1}{4} \bigg( \bar{\nabla}_i - \ftb{1}{2} ({\cal D}^p {\cal E}_{pi} + 4 \bar{\cal D}_i d ) \bigg) \bar{\eta}^i \label{gtform2} \, ,
\ee
where $\eta_i \equiv - \tilde{\xi}_i + {\cal E}_{ij} \xi^j$ and $\bar{\eta}_i \equiv \tilde{\xi}_i + {\cal E}_{ji} \xi^j$. 
Note that in the limit of no $\tilde{x}$ dependence, these gauge transformations reduce to the gauge transformations \eqref{gaugetrfgr} in General Relativity. The statement of gauge invariance is as follows:
\bq \label{gaugeinv}
S[ {\cal E}+ \delta {\cal E}, d + \delta d] = S[{\cal E}, d] \, ,
\eq
where $\delta {\cal E}$ and $\delta d$ are given by \eqref{gtform1} and \eqref{gtform2}, respectively. \\
The variation of the double field theory action \eqref{action} is
\bq
\ba{lcl}
\delta S &=& \delta_{\cal E} S + \delta_d S \ = \displaystyle\int dx d \tilde{x}  \ e^{-2d} \delta {\cal E}_{pq} {\cal K}^{pq} \ + \ \displaystyle\int dx d \tilde{x}  \ e^{-2d} ( -2 \delta d) {\cal R} \\ [2.5ex]
&=&  \displaystyle\int dx d \tilde{x}  \ e^{-2d} \Bigg[  \bigg\{ \bar{\nabla}_q \eta_p {\cal K}^{pq} + \ftb{1}{2} \bigg( \nabla_p - \ftb{1}{2} (\bar{\cal D}^q {\cal E}_{pq} + 4 {\cal D}_p d ) \bigg) \eta^p  {\cal R} \bigg\} \\
&& + \bigg\{ \nabla_p \bar{\eta}_q {\cal K}^{pq} + \ftb{1}{2} \bigg( \bar{\nabla_q} - \ftb{1}{2} ({\cal D}^p {\cal E}_{pq} + 4 \bar{\cal D}_q d ) \bigg) \bar{\eta}^q {\cal R} \bigg\} \Bigg] .
\ea
\eq
Then using the integration by parts we obtain the following:
\bq \label{bianchiprev}
\ba{lcl}
\delta S &=& \displaystyle\int dx d \tilde{x}  \ e^{-2d} \Bigg[ -  \eta^p \bigg\{ \Big( \bar{\nabla}^q - \ftb{1}{2} g^{qs} ({\cal D}^r {\cal E}_{rs} + 4 \bar{\cal D}_s d) \Big)  {\cal K}_{pq}  + \ftb{1}{2} {\cal D}_p {\cal R}  \bigg\} \\
&& - \bar{\eta}^q  \bigg\{ \Big( {\nabla}^p - \ftb{1}{2} g^{ps} (\bar{\cal D}^r {\cal E}_{sr} + 4 {\cal D}_s d) \Big) {\cal K}_{pq} + \ftb{1}{2} \bar{\cal D}_q {\cal R} \bigg\} \Bigg] \, .
\ea
\eq
From the gauge invariance principle \eqref{gaugeinv}, the above variation should vanish for arbitrary gauge parameters $\eta$ and $\bar{\eta}$. Since $\eta$ and $\bar{\eta}$ are independent, this yields the two independent Bianchi identities given in \eqref{final_bi1} and \eqref{final_bi2}, which have forms similar to the contracted Bianchi identity in Appendix \ref{usualgr}.
\subsection{The limit of no $\tilde{x}$ dependence}
Here we first compare our equations of motion in the limit of no $\tilde{x}$ dependence with those in a conventional GR setup. In this limit, the calligraphic derivatives defined in \eqref{calderdef} simplify as
\bq
{\cal D}_i = \bar{\cal D}_i = \partial_i \, .
\eq
The equation of motion for $\cal E$ in the limit of no $\tilde{x}$ dependence is obtained by taking \eqref{kpqlimit} to be zero.
It is easy to see that this equation of motion is covariant under diffeomorphisms. Since the terms in the first square bracket are symmetric and the terms in the second square bracket are antisymmetric, the first and second square brackets should vanish separately.\\
\\
In this limit the $O(D,D)$ Ricci-like scalar is given in~\cite{Hohm:2010jy} as follows:
\bq
{\cal R} \Big|_{\tilde{\partial}=0} = R + 4(\nabla^i \nabla_i \phi - (\partial \phi)^2) - \ftb{1}{12}H^2 \, .
\eq
Thus the equations of motion of double field theory in the limit of no $\tilde{x}$ dependence are summarized as
\bq
\ba{rcl} \label{eom}
R_{ij}  - \ftb{1}{4} {H_{i}}^{ p q} H_{j p q}  + 2 \nabla_i \nabla_j \phi &=& 0 \, , \\[1.5ex]
\ftb{1}{2} \nabla^p H_{p i j} - H_{p i j} \nabla^p \phi &=& 0 \, , \\ [1.5ex]
R + 4(\nabla^i \nabla_i \phi - (\partial \phi)^2) - \ftb{1}{12}H^2 &=&0 \, .
\ea
\eq
In this limit, it can be directly seen from Appendix \ref{usualgr} that the second and third equations of motion given in \eqref{eom} are exactly the same as those for the antisymmetric tensor $b_{ij}$ and the scalar $\phi$ of General Relativity, respectively. However, the first equation in \eqref{eom} is different from the equation of motion for $g_{ij}$ in Appendix \ref{usualgr}. This difference originates from the definition of a density in double field theory as compared to General Relativity. Since $e^{-2d} \equiv \sqrt{-g} e^{-2 \phi}$, the $g_{ij}$-variation gives some extra terms in General Relativity due to the $\sqrt{-g}$ factor in front of the exponential term. Due to this fact, a linear combination of the equations of motion for $g_{ij}$ and $\phi$ yields exactly our first equation in \eqref{eom} as follows:
\bq
\ba{l}
 \bigg[ R_{i j} - \ftb{1}{2} g_{i j} R +2 \Big( -g_{i j} \nabla^2 \phi + \nabla_{i} \nabla_{j} \phi + g_{i j} (\partial \phi)^2 \Big)  - \ftb{1}{4} \Big( {H_i}^{p q} H_{j p q} - \ftb{1}{6} g_{i j} H^2 \Big) \bigg]  \\  [2.0ex]
 + \ftb{1}{2} g_{i j}  \bigg[ R+4 \Big( \nabla^2 \phi - (\partial \phi)^2 \Big) - \ftb{1}{12} H^2  \bigg] 
 = R_{i j} + 2 \nabla_{i} \nabla_{j} \phi - \ftb{1}{4}  {H_i}^{p q} H_{j p q} = 0 \, .
\ea
\eq
\\
Now we can take this limit for the Bianchi identities of $O(D,D)$ tensors \eqref{final_bi1} and \eqref{final_bi2}.
With no $\tilde{x}$-dependence these Bianchi identities respectively reduce to
\bq \label{Bianchi}
 \ftb{1}{2} \bigg[ \nabla^p \nabla^s H_{spq}  + \Big( \ftb{1}{2}  {H_p}^{r s} \nabla^p H_{q r s} - \ftb{1}{12} \nabla_q H^2 \Big)  \bigg] \\ + \bigg( -\nabla^p R_{pq} + \ftb{1}{2} \nabla_q R \bigg) =0 \, ,
\eq
\bq \label{Bianchi2}
 \ftb{1}{2} \bigg[ \nabla^q \nabla^s H_{spq}  + \Big( \ftb{1}{2}  {H_q}^{ r s} \nabla^q H_{p  r s} - \ftb{1}{12} \nabla_p H^2 \Big)  \bigg] \\ + \bigg( -\nabla^q R_{pq} + \ftb{1}{2} \nabla_p R \bigg) =0 \, .
\eq
The first terms in \eqref{Bianchi} and \eqref{Bianchi2} vanish from the symmetries of the Ricci tensor and the Riemann curvature tensor as follows:
\bq \label{JacobiRiemann}
\ba{lcl}
\nabla^i \nabla^j H_{ijk} &=&  \ftb{1}{2} [ \nabla^i , \nabla^j ] H_{ijk} = \ftb{1}{2} [\nabla_{i} , \nabla_{j}] {H^{i j}}_k \\ [2.0ex]
&=&  \ftb{1}{2} \bigg[ {R^i}_{l i j} {H^{l j}}_k +{R^{j}}_{l i j} {H^{i l}}_k  +R_{k l i j} H^{i j l }  \bigg] = 0 \, ,
\ea
\eq
where the first term and the second term in the last line vanish due to the symmetric property of Ricci tensor, while the last term in the last line vanishes due to the Bianchi identity of the Riemann curvature tensor, $R_{l [i j k]} =0$. \\
\\
The second and third term in \eqref{Bianchi} cancel each other from the Bianchi identity of $H$-field as follows:
\bq \label{JacobiH}
0 = H^{p r s} \nabla_{[p} H_{r s q]} = \ftb{3}{4} \bigg(H^{p r s} \nabla_{p} H_{q r s} - \ft{1}{6} \nabla_q H^2 \bigg)  \, ,
\eq
where the square bracket for indices means antisymmetrization. The same cancellation applies to \eqref{Bianchi2}.\\
The Bianchi identities \eqref{Bianchi} and \eqref{Bianchi2} in the limit of no $\tilde{x}$ dependence then reduce to
\bq
 \nabla^i R_{ij} - \ftb{1}{2} \nabla_j R  = 0 \, ,
\eq
which is the contracted Bianchi identity in General Relativity.


\section{Gauge transformation of ${\cal K}_{pq}$} \label{gaugetr section}
The $O(D,D)$ tensor ${\cal K}_{pq}$ derived in \eqref{kpqlimit} has unusual gauge transformation properties. In this section we again use the gauge invariance of the double field theory action \eqref{action} to obtain the gauge transformation of ${\cal K}_{pq}$. Then we offer some remarks on the gauge transformation of ${\cal K}_{pq}$. The gauge transformation of ${\cal K}_{pq}$ is also discussed in the next section using the generalized metric formulation.
\subsection{Gauge transformation from the gauge invariance principle}
Consider the gauge transformation with gauge parameters $\xi$ and $\tilde{\xi}$. From~\cite{Hohm:2010jy} we have the following gauge transformation of ${\cal E}$:
\bq
\delta_{G} {\cal E}_{ij} = \partial_i \tilde{\xi}_j - \partial_j \tilde{\xi}_i + {\cal L}_{\xi} {\cal E}_{ij} + {\cal L}_{\tilde{\xi}} {\cal E}_{ij}  + {\cal E}_{ik} (\tilde{\partial}^q \xi^k - \tilde{\partial}^k \xi^q) {\cal E}_{qj} \, ,
\eq
which is equivalent to the first equation in \eqref{firstgaugetransformation}. Here $\delta_G$ is used to distinguish the gauge transformation from an arbitrary variation $\delta$, which appears later. Now define $T^{kq} \equiv \tilde{\partial}^q \xi^k - \tilde{\partial}^k \xi^q$ such that the above gauge transformation of $\cal E$ can be written as
\bq \label{gtcale}
\delta_{G} {\cal E}_{ij} = \partial_i \tilde{\xi}_j - \partial_j \tilde{\xi}_i + {\cal L}_{\xi} {\cal E}_{ij} + {\cal L}_{\tilde{\xi}} {\cal E}_{ij}  + {\cal E}_{ik} T^{kq} {\cal E}_{qj} \, .
\eq
Consider an infinitesimal variation  $\delta {\cal E}$ for some background field ${\cal E}$ such that ${\cal E}' = {\cal E} + \delta {\cal E}$. The statement of gauge invariance implies,
\bq \label{gtgaugeinv}
\delta_G \Big(S[{\cal E}' , d] - S[{\cal E} , d] \Big) = 0 \, ,
\eq
which can be rewritten as follows:
\bq \label{gtgaugeinv2}
 \delta_G \bigg[\displaystyle\int dx d \tilde{x}  \ e^{-2d} \delta {\cal E}_{pq} {\cal K}^{pq}  \bigg] 
 = \displaystyle\int dx d \tilde{x} \Bigg[  \delta_G \Big( e^{-2d} \Big) \delta {\cal E}_{pq} {\cal K}^{pq}  + e^{-2d} \delta_G \Big( \delta {\cal E}_{pq} {\cal K}^{pq} \Big) \Bigg] = 0  \, .
\eq
It can be easily seen that \eqref{gtgaugeinv2} holds if $\delta_G ( \delta {\cal E}_{pq} {\cal K}^{pq}) = \xi^M \partial_M  ( \delta {\cal E}_{pq} {\cal K}^{pq} )$ since $e^{-2d}$ transforms as a density under gauge transformation. Since the gauge transformation $\delta_G$ is a linear operation, we have
\bq \label{lingauge}
\delta_G \Big( \delta {\cal E}_{pq} {\cal K}^{pq} \Big) = \delta_G \Big( \delta {\cal E}_{pq} \Big) {\cal K}^{pq} + \delta {\cal E}_{pq} \delta_G {\cal K}^{pq} \, .
\eq
Then one can write $\delta_G {\cal K}^{pq}$ as
\bq 
\delta_G {\cal K}^{pq} = \hat{\delta}  {\cal K}^{pq} + \theta^{pq} \, ,
\eq
where $ \hat{\delta}  {\cal K}^{pq} $ is the part of the transformation that gives the desired result for $\delta_G ( \delta {\cal E}_{pq} {\cal K}^{pq})$:
\bq
 \delta_G \Big( \delta {\cal E}_{pq} \Big) {\cal K}^{pq} + \delta {\cal E}_{pq}  \hat{\delta} {\cal K}^{pq}    = \xi^M \partial_M  \Big( \delta {\cal E}_{pq} {\cal K}^{pq} \Big) \, .
\eq
Using this decomposition, \eqref{lingauge} gives
\bq 
\delta_G \Big( \delta {\cal E}_{pq} {\cal K}^{pq} \Big) = \xi^M \partial_M  \Big( \delta {\cal E}_{pq} {\cal K}^{pq} \Big) + \delta {\cal E}_{pq} \theta^{pq} \, ,
\eq
where $\theta^{pq}$, 
on account of \eqref{gtgaugeinv2}, 
 should satisfy
\bq \label{thetagcon}
\displaystyle\int dx d \tilde{x}  \, e^{-2d} \delta {\cal E}_{pq} 
\theta^{pq} = 0 \, .
\eq
Since $\delta {\cal E}$ is arbitrary $\theta^{pq}$ must vanish. This finally shows  
that the gauge transformation of ${\cal K}^{pq}$ is given~by
\bq 
\delta {\cal E}_{pq} \delta_G  {\cal K}^{pq}    = \xi^M \partial_M  \Big( \delta {\cal E}_{pq} {\cal K}^{pq} \Big) -  \delta_G \Big( \delta {\cal E}_{pq} \Big) {\cal K}^{pq} \, .
\eq
From \eqref{gtcale} the gauge transformation of $\delta {\cal E}_{pq}$ is
\bq
\delta_G \Big( \delta {\cal E}_{pq} \Big) = \delta_G \Big(  {\cal E}'_{pq} - {\cal E}_{pq} \Big) = {\cal L}_{\xi} \delta {\cal E}_{pq} + {\cal L}_{\tilde{\xi}}  \delta {\cal E}_{pq}  + \delta {\cal E}_{pk} T^{kl} {\cal E}_{lq} + {\cal E}_{pk} T^{kl}  \delta {\cal E}_{lq} \, .
\eq
Thus one can write
\bq
\delta {\cal E}_{pq} \delta_G  {\cal K}^{pq}    =  \xi^M \partial_M  \Big( \delta {\cal E}_{pq} {\cal K}^{pq} \Big)  - \Bigg[ {\cal L}_{\xi} \delta {\cal E}_{pq} + {\cal L}_{\tilde{\xi}}  \delta {\cal E}_{pq}  + \delta {\cal E}_{pk} T^{kl} {\cal E}_{lq} + {\cal E}_{pk} T^{kl}  \delta {\cal E}_{lq} \Bigg] {\cal K}^{pq}  \, .
\eq
From $ \xi^M \partial_M ( \delta {\cal E}_{pq} {\cal K}^{pq}) = {\cal L}_{\xi} ( \delta {\cal E}_{pq} {\cal K}^{pq} ) + {\cal L}_{\tilde{\xi}} ( \delta {\cal E}_{pq} {\cal K}^{pq} )$, one obtains
\bq
\delta {\cal E}_{pq} \delta_G  {\cal K}^{pq}  = \delta {\cal E}_{pq} \Bigg[ {\cal L}_{\xi} {\cal K}^{pq} + {\cal L}_{\tilde{\xi}}   {\cal K}^{pq} - T^{ql} {\cal E}_{lk} {\cal K}^{pk} - T^{kp} {\cal E}_{lk} {\cal K}^{lq} \Bigg]  \, .
\eq
Then we find the gauge transformation of ${\cal K}^{pq}$ as follows:
\bq
\delta_G  {\cal K}^{pq}  = {\cal L}_{\xi} {\cal K}^{pq} + {\cal L}_{\tilde{\xi}}   {\cal K}^{pq} - T^{ql} {\cal E}_{lk} {\cal K}^{pk} - T^{kp} {\cal E}_{lk} {\cal K}^{lq} \, .
\eq
Given the gauge transformation of ${\cal K}^{pq}$, it is straightforward to derive the gauge transformation of ${\cal K}_{pq}$ once we know the gauge transformation of $g_{ij}$. The gauge transformation of $g_{ij} = \ft{1}{2} ({\cal E}_{ij} + {\cal E}_{ji})$ can be directly calculated from the gauge transformation of ${\cal E}_{ij}$ \eqref{gtcale}, as follows:
\bq
\delta_G g_{ij}  = {\cal L}_{\xi} g_{ij} +  {\cal L}_{\tilde{\xi}} g_{ij} + \ftb{1}{2}  \Big( {\cal E}_{ik} T^{kq} {\cal E}_{qj} - {\cal E}_{ki} T^{kq} {\cal E}_{jq} \Big) \, ,
\eq
where the property $T^{kq} = - T^{qk}$ is used. Finally we find the gauge transformation of ${\cal K}_{pq}$ from the definition ${\cal K}_{pq } = g_{pi} {\cal K}^{ij} g_{jq}$. The result is given by
\bq
\ba{lcl}
\delta_G  {\cal K}_{pq } &=&  \delta_G  g_{pi} {\cal K}^{ij } g_{jq} +  g_{pi} \delta_G  {\cal K}^{ij }  g_{jq} + g_{pi}  {\cal K}^{ij } \delta_G   g_{jq}  \\ [1.0ex]
&=&  {\cal L}_{\xi} {\cal K}_{pq} + {\cal L}_{\tilde{\xi}}   {\cal K}_{pq} + {\cal E}_{pk} T^{kl} {\cal K}_{lq} + {\cal E}_{lq} T^{kl} {\cal K}_{pk}   \, .
\ea
\eq
Thus the full gauge transformation of ${\cal K}_{pq}$ is given by
 \bq \label{fullgt_kpq}
\delta_G {\cal K}_{pq} = {\cal L}_{\xi} {\cal K}_{pq} + {\cal L}_{\tilde{\xi}} {\cal K}_{pq} + {\cal E}_{pk} (\tilde{\partial}^l \xi^k - \tilde{\partial}^k \xi^l ) {\cal K}_{lq} + {\cal E}_{lq} (\tilde{\partial}^l \xi^k - \tilde{\partial}^k \xi^l ) {\cal K}_{pk} \, .
\eq
We explicitly checked the gauge transformation of ${\cal K}_{pq}$ with both perturbative calculations and non-perturbative calculations (along the lines of section 4.2 of~\cite{Hohm:2010jy}). From the perturbative calculation, we obtained the same result \eqref{fullgt_kpq} up to second order in gauge parameters and fields. Non-perturbative calculation provided the verification that a few terms in \eqref{finalkpq} have the gauge transformation property given by \eqref{fullgt_kpq}.
\subsection{Properties of gauge transformation of ${\cal K}_{pq}$}
In the limit of no $\tilde{x}$ dependence, the gauge transformation \eqref{fullgt_kpq} of ${\cal K}_{pq}$ reduces to
\bq
\delta_G {\cal K}_{pq} \bigg|_{\tilde{\partial} =0} = {\cal L}_{\xi} {\cal K}_{pq} \bigg|_{\tilde{\partial} =0} \, .
\eq
Hence, as expected ${\cal K}_{pq}$ reduces to a covariant tensor under diffeomorphisms.
\\
Now \eqref{fullgt_kpq} can be written in terms of $O(D,D)$ covariant derivatives and parameters. $O(D,D)$ covariant gauge parameters are defined as follows:
\be
\eta_i \equiv - \tilde{\xi}_i + {\cal E}_{ij} \xi^j  & , & \bar{\eta}_i \equiv \tilde{\xi}_i + {\cal E}_{ji} \xi^j \, ,
\ee
where $\eta$ transforms covariantly under the $O(D,D)$ transformation with unbarred index while $\bar{\eta}$ transforms covariantly with barred index. The gauge transformation \eqref{fullgt_kpq} of ${\cal K}_{pq}$ can be rewritten in terms of $\eta$, $\bar{\eta}$, $\nabla$, and $\bar{\nabla}$ as
\bq \label{bi_gt}
\ba{lcl}
\delta_G {\cal K}_{pq} &=& \ftb{1}{2} (\eta \cdot \nabla + \bar{\eta} \cdot \bar{\nabla}) {\cal K}_{pq} + \ftb{1}{2} (\bar{\nabla}_q \bar{\eta}^k - \bar{\nabla}^k \bar{\eta}_q) {\cal K}_{pk} + \ftb{1}{2} (\nabla_p \eta^k - \nabla^k \eta_p) {\cal K}_{kq} \\ [2.0ex]
&&+ \ \underline{\ftb{1}{2} g^{kl} (\bar{\nabla}_q \eta_l + \nabla_l \bar{\eta}_q)  {\cal K}_{pk}  + \ftb{1}{2} g^{kl} (\nabla_p \bar{\eta}_l + \bar{\nabla}_l \eta_p) {\cal K}_{kq} } \ ,
\ea
\eq
where underlined terms violate the $O(D,D)$ covariance hence $\delta_G {\cal K}_{pq}$ is not $O(D,D)$ covariant. \\
From the gauge transformation \eqref{gtform2} of the dilaton, one can manifestly see that $\delta d$ is an $O(D,D)$ scalar. Thus terms in $\delta_G {\cal K}_{pq}$ related to the gauge transformation of $d$ are $O(D,D)$ covariant. These terms are included in the first line of \eqref{bi_gt}. Then underlined terms in \eqref{bi_gt}, which are not $O(D,D)$ covariant, are from the gauge transformation of $\cal E$, which is given in \eqref{gtform1}. Thus $O(D,D)$ non-covariant terms in \eqref{bi_gt} can be written as follows:
\bq
\delta_G {\cal K}_{pq} = \bigg( O(D,D) \text{ covariant terms} \bigg)+ \underline{\ftb{1}{2} g^{kl} \delta {\cal E}_{lq}  {\cal K}_{pk}  + \ftb{1}{2} g^{kl} \delta {\cal E}_{pl} {\cal K}_{kq} } \, ,
\eq
which is consistent with the theorem in section \ref{altodd}.
\section{Generalized metric formulation} \label{genmetsec}
The work in~\cite{Hohm:2010pp} describes the background independent double field theory~\cite{Hohm:2010jy} using the generalized metric $\cal H$. The action is defined in \eqref{genmetaction} and its $\cal H$-variation ${\cal K}_{MN}$ is given in \eqref{kmn}. By carefully varying the action with respect to $\cal H$, preserving the constraint for $\cal H$, ${\cal R}_{MN}$ is defined by \eqref{ma_form} whose vanishing is the equation of motion for the field ${\cal H}^{MN}$. Since this generalized metric formulation is just a different description of the work in~\cite{Hohm:2010jy} with the same constraint, ${\cal K}_{MN}$ and ${\cal R}_{MN}$ must be closely related to our ${\cal K}_{pq}$. This relation is more clear in the vielbein formulation of Siegel~\cite{Siegel:1993th, Siegel:1993xq} as elaborated in~\cite{Hohm:2010pp}. In this formulation the frame field $e^M{}_A$ was introduced to define frame components of a tensor with the usual vector indices $M, N, \cdots$. By fixing the $GL(D, \mathbb{R}) \times GL(D, \mathbb{R})$ symmetry of the frame field, the generalized metric formulation with $\cal H$ can be written in terms of ${\cal E}_{ij}$.\\
\\
We will first start with the gauge transformation properties of ${\cal K}_{MN}$ and ${\cal R}_{MN}$ and show that ${\cal R}_{MN}$ is a generalized tensor. Then we will prove the equivalence (up to a sign) between ${\cal R}_{a \bar{b}} \equiv e_a^M {\cal R}_{MN} e_{\bar{b}}^N$ and our ${\cal K}_{a\bar{b}}$ defined in \eqref{finalkpq} after gauge-fixing the $GL(D, \mathbb{R}) \times GL(D, \mathbb{R})$ symmetry. Finally we give another derivation of the gauge transformation of ${\cal K}_{pq}$ using the gauge transformation properties of a generalized tensor.
\subsection{Proof that ${\cal R}_{MN}$ is a generalized tensor from gauge invariance principle }
Here we first investigate the gauge transformation property of ${\cal K}_{MN}$ then prove that ${\cal R}_{MN}$ is a generalized tensor from this property.
Consider background fields ${\cal H}'$ and ${\cal H}$ which are related by ${\cal H}' = {\cal H} + \delta {\cal H}$. The gauge invariance of the action \eqref{genmetaction} implies
\bq \label{genmetgaugeinv}
\delta_{\xi} \bigg[ S[{\cal H}', d ]  - S[{\cal H}, d ] \bigg] = \delta_{\xi}  \displaystyle\int dx d \tilde{x}  \ e^{-2d} \delta {\cal H}^{MN} {\cal K}_{MN}  = 0 \, ,
\eq
where $\delta_{\xi}$ denotes gauge transformations in this formulation. \\
Now the gauge transformation property of ${\cal H}$ is given by the following:
\bq
\delta_{\xi} {\cal H}^{MN} = \hat{\cal L}_{\xi} {\cal H}^{MN}  \Longrightarrow  \delta_{\xi} \Big( \delta {\cal H}^{MN} \Big) =  \hat{\cal L}_{\xi}  \Big( \delta {\cal H}^{MN} \Big) \, ,
\eq
where a generalized Lie derivative $\hat{\cal L}_{\xi}$ is defined in \eqref{genlied}.  \\
Consider an object $W$ in the generalized metric formulation. The gauge transformation of $W$ can be decomposed into two parts: one corresponding to a generalized Lie derivative and the other being additional terms. This can be written as
\bq \label{decomp}
\delta_{\xi} W = \hat{\cal L}_{\xi} W +  \Delta_{\xi} W \, ,
\eq
so that $W$ is a generalized tensor if and only if $\Delta_{\xi} W$ vanishes. \\
\\
Using the same decomposition, $\delta_{\xi} {\cal K}_{MN}= \hat{\cal L}_{\xi} {\cal K}_{MN} + \big( \Delta_{\xi} {\cal K} \big)_{MN} \,$. Then \eqref{genmetgaugeinv} becomes
\bq \label{deltaxicon}
\ba{lcl}
0 &=&   \displaystyle\int dx d \tilde{x} \bigg[  \delta_{\xi} \Big( e^{-2d} \Big) \delta {\cal H}^{MN} {\cal K}_{MN} +   e^{-2d} \delta_{\xi} \Big(  \delta {\cal H}^{MN} \Big) {\cal K}_{MN} + e^{-2d}  \delta {\cal H}^{MN}  \delta_{\xi} \Big( {\cal K}_{MN}  \Big) \bigg] \\ [2.0ex]
&=&  \displaystyle\int dx d \tilde{x} \bigg[  \delta_{\xi} \Big( e^{-2d} \Big) \delta {\cal H}^{MN} {\cal K}_{MN} + e^{-2d} \hat{\cal L}_{\xi} \bigg(  \delta {\cal H}^{MN} {\cal K}_{MN}  \bigg) +  e^{-2d}  \delta {\cal H}^{MN}  \big( \Delta_{\xi} {\cal K} \big)_{MN} \bigg] \\ [2.0ex]
&=&  \displaystyle\int dx d \tilde{x}  e^{-2d}  \delta {\cal H}^{MN}  \big( \Delta_{\xi} {\cal K} \big)_{MN}  \, ,
\ea
\eq
where $ \hat{\cal L}_{\xi} (  \delta {\cal H}^{MN} {\cal K}_{MN}  ) = \xi^L \partial_L  (  \delta {\cal H}^{MN} {\cal K}_{MN}  )$ and the gauge transformation property of $e^{-2d}$ as a density are used for the first two terms in the second line of \eqref{deltaxicon} to vanish. The constraint for $\delta {\cal H}$ is given by the equation (4.52) of~\cite{Hohm:2010pp}:
\bq
\delta {\cal H}^{MN} = - S^M{}_P \delta {\cal H}^{PQ} S^N{}_Q \, .
\eq
From the definition of the matrix $S^M{}_N \equiv {\cal H}^{M}{}_{N}$, by lowering indices using $\eta_{KM}$ and $\eta_{LN}$, one obtains
\bq
\delta {\cal H}^{MN} \Big( \eta_{KM} \eta_{LN} + {\cal H}_{KM} {\cal H}_{LN} \Big) = 0 \, .
\eq
Thus, to satisfy \eqref{deltaxicon}, $ ( \Delta_{\xi} {\cal K} )_{MN}$ must take the following form:
\bq 
\big( \Delta_{\xi} {\cal K} \big)_{MN} =  \Big( \eta_{KM} \eta_{LN} + {\cal H}_{KM}  {\cal H}_{LN} \Big) {\cal T}^{KL}  = {\cal T}_{MN} + {S^K}_{M} {\cal T}_{KL} {S^L}_{N} \, ,
\eq
for some tensor $\cal T$. In the matrix form, this can be rewritten as
\bq \label{kmnform}
\big( \Delta_{\xi} {\cal K} \big) = {\cal T} + S^t {\cal T} S \, .
\eq
From this argument, ${\cal K}_{MN}$ does not necessarily need to be a generalized tensor. With a detailed calculation, we have verified that $\Delta_{\xi} {\cal K}$ takes the form \eqref{kmnform} with $\cal T$ given by
\bq
\ba{lcl}
{\cal T}_{MN} &=& - \ftb{1}{4} \partial_M \partial_N (\partial \cdot \xi) + \ftb{1}{2}  \partial_M \partial_N \xi \cdot \partial d 
+ \ftb{1}{4}  (\partial_M {{\cal H}^L}_P ) {{\cal H}^K}_N \partial_L (\partial_K \xi^P - \partial^P \xi_K)  \\ [2.0ex]
&&- \ \ftb{1}{2} {{\cal H}^P}_M  \partial_L \partial_P \xi \cdot \partial  {{\cal H}^L}_N  + \bigg\{ M \leftrightarrow N \bigg\} \, .
\ea
\eq
Thus $\Delta_{\xi} {\cal K}$ has the claimed form above. \\
\\
Now we prove that ${\cal R}_{MN}$ is a generalized tensor. This can be shown by verifying the following:
\bq
\delta_{\xi} {\cal R}_{MN} = \hat{\cal L}_{\xi} {\cal R}_{MN}  \quad \Longleftrightarrow \quad \Delta_{\xi} {\cal R}_{MN} =0 \, ,
\eq
where $\Delta_{\xi} {\cal R}_{MN} $ is defined by the decomposition \eqref{decomp}. Since $\delta_{\xi} S^M{}_N = \hat{\cal L}_{\xi} S^M{}_N$, $\Delta_{\xi} {\cal R}_{MN}$ is given by
\bq
\Delta_{\xi} {\cal R} = \ftb{1}{4} (1- S^t) \Delta_{\xi} {\cal K} (1+S) + \ftb{1}{4} (1+ S^t) \Delta_{\xi} {\cal K} (1-S) \, ,
\eq
from the definition of ${\cal R}_{MN}$ \eqref{ma_form}. Using $S^2 = 1$ and the form of $\Delta_{\xi} {\cal K}$ \eqref{kmnform}, we obtain the following result:
\bq
\Delta_{\xi} {\cal R} = \ftb{1}{4} (1-S^t) \Big[{\cal T} + S^t {\cal T} S \Big] (1+S) + \ftb{1}{4} (1+S^t) \Big[{\cal T} + S^t {\cal T}S \Big] (1-S) =0 \, ,
\eq
which proves the statement that ${\cal R}_{MN}$ is a generalized tensor.
\subsection{Explicit check of equivalence between ${\cal R}_{a \bar{b}}$ and ${\cal K}_{pq}$} \label{checkrandk}
The frame component ${\cal R}_{a \bar{b}}$ of ${\cal R}_{MN}$ is defined by (5.62) of~\cite{Hohm:2010pp}:
\bq
{\cal R}_{a \bar{b}} = {e^M}_{a} {\cal K}_{MN} {e^N}_{\bar{b}} \, .
\eq
We will show that ${\cal R}_{a \bar{b}} = - {\cal K}_{a \bar{b}}$, where ${\cal K}_{a \bar{b}}$ is defined in \eqref{finalkpq}, once the $GL(D, \mathbb{R} ) \times GL(D, \mathbb{R} )$ symmetry is gauge-fixed as follows:
\bq
e^{M}{}_A \ = \ \begin{pmatrix} e_{i \bar{a}} &  e_{i a } \\ e^i{}_{\bar{a}} & e^i{}_a \end{pmatrix} 
   \ = \ \begin{pmatrix} {\cal E}_{i\bar{a}}   &  -{\cal E}_{ai}  \\ \delta^{i}{}_{\bar{a}}  & \delta^{i}{}_{a}  \end{pmatrix}\;.
\eq
This gauge fixing enables us to compare the results in vielbein formalism with the results written in terms of ${\cal E}$. Here we will use the identities below. From (4.16) of~\cite{Hohm:2010pp}, the generalized metric $\cal H$ can be rewritten as follows:
\bq
{\cal H}^{MN} = e^M{}_i e^N{}_j g^{ij} - \eta^{MN} \, .
\eq
The calligraphic derivatives in the double field theory are given by
\bq
{\cal D}_a = e^M{}_a \partial_M \ , \quad \bar{\cal D}_a = e^M{}_{\bar{a}} \partial_M \, ,
\eq
which is from (5.58) of~\cite{Hohm:2010pp}. \\
\\
Now using the above two identities with some amount of calculation, we obtain the following result for all terms containing the dilaton $d$ in \eqref{kmn}:
\bq
\ba{l}
{e^M}_{a} \bigg[ 2 \partial_M \partial_N d + \ftb{1}{2} (\partial_L d) ({\cal H}^{LK} \partial_K {\cal H}_{MN}) -  (\partial_L d) ({\cal H}^{KL} \partial_{(M} {\cal H}_{N)K} + {{\cal H}^K}_{(M} \partial_K {{\cal H}^L}_{N)} ) \bigg] {e^N}_{\bar{b}} \\ [2.0ex]
= \nabla_a \bar{\cal D}_b d + \bar{\nabla}_b {\cal D}_a d \, .
\ea
\eq
These coincide (up to a sign) with the dilaton terms in \eqref{finalkpq}. Other terms in \eqref{kmn} are calculated term by term, which yields the following results:
\bq
\ftb{1}{8}  {e^M}_{a} \partial_M {\cal H}^{KL} \partial_N {\cal H}_{KL}  {e^N}_{\bar{b}} = - \ftb{1}{4} g^{kl} g^{mn} {\cal D}_a {\cal E}_{km} \bar{\cal D}_b {\cal E}_{ln}  \, ,
\eq
\bq
- \ftb{1}{4}  {e^M}_{a} \partial_L ({\cal H}^{LK} \partial_K {\cal H}_{MN}) {e^N}_{\bar{b}}  = - \ftb{1}{2} \bar{\cal D}^k \bar{\cal D}_k {\cal E}_{ab} + \ftb{1}{2} g^{ij} {\cal D}^k {\cal E}_{ai} \bar{\cal D}_{jb} + \ftb{1}{4} ({\cal D}^k {\cal E}_{kl} + \bar{\cal D}^k {\cal E}_{lk} ) \bar{\cal D}^l {\cal E}_{ab} \, ,
\eq
\bq
- \ftb{1}{2}  {e^M}_{a}  \partial_{(M} {\cal H}^{KL} \partial_L {\cal H}_{N)K} {e^N}_{\bar{b}} = \ftb {1}{4} g^{kl} ({\cal D}_a {\cal E}_{lq} \bar{\cal D}^q {\cal E}_{kb} + \bar{\cal D}_b {\cal E}_{ql} {\cal D}^q {\cal E}_{ak} ) \, ,
\eq
\bq
\ftb{1}{2}  {e^M}_{a} \partial_L {\cal H}^{KL} \partial_{(M} {\cal H}_{N)K} {e^N}_{\bar{b}} = -\ftb{1}{4} g^{jl} (\bar{\cal D}^p {\cal E}_{lp} {\cal D}_a {\cal E}_{jb} + {\cal D}^p {\cal E}_{pl} \bar{\cal D}_b {\cal E}_{aj}) \, ,
\eq
\bq
\ftb{1}{4} {e^M}_{a} {\cal H}^{KL} \partial_L \partial_M {\cal H}_{NK} {e^N}_{\bar{b}} = \ftb{1}{4} {\cal D}_a {\cal D}^p {\cal E}_{pb} + \ftb{1}{4} g^{jl} \bar{\cal D}^p {\cal E}_{jb} {\cal D}_a {\cal E}_{lp} + \ftb{1}{8} g^{jl} {\cal D}_a {\cal E}_{jb} \bigg( \bar{\cal D}^p {\cal E}_{lp} -  {\cal D}^p {\cal E}_{pl} \bigg) \, ,
\eq
\bq
\ftb{1}{4} {e^M}_{a} {\cal H}^{KL} \partial_L \partial_N {\cal H}_{MK} {e^N}_{\bar{b}} = \ftb{1}{4} \bar{\cal D}_b \bar{\cal D}^p {\cal E}_{ap} + \ftb{1}{4} g^{jl} {\cal D}^p {\cal E}_{aj} \bar{\cal D}_b {\cal E}_{pl} - \ftb{1}{8} g^{jl} \bar{\cal D}_b {\cal E}_{aj} \bigg( \bar{\cal D}^p {\cal E}_{lp} -  {\cal D}^p {\cal E}_{pl} \bigg) \, ,
\eq
\bq
\ftb{1}{2} {e^M}_{a} \partial_L {{\cal H}^K}_{(M} \partial_K {{\cal H}^L}_{N)} {e^N}_{\bar{b}} = -\ftb{1}{2} {\cal D}^p {\cal E}_{aq} \bar{\cal D}^q {\cal E}_{pb} \, ,
\eq
\bq
\ftb{1}{4}  {e^M}_{a}  {{\cal H}^K}_{M} \partial_L \partial_K {{\cal H}^L}_{N} {e^N}_{\bar{b}} = \ftb{1}{4} \bigg[ {\cal D}_a {\cal D}^p {\cal E}_{pb} - \ftb{1}{2} g^{jl} {\cal D}_a {\cal E}_{jb} \Big( {\cal D}^p {\cal E}_{pl} + \bar{\cal D}^p {\cal E}_{lp}  \Big) \bigg] \, ,
\eq
\bq
\ftb{1}{4}  {e^M}_{a}  {{\cal H}^K}_{N} \partial_L \partial_K {{\cal H}^L}_{M} {e^N}_{\bar{b}} = \ftb{1}{4} \bigg[ \bar{\cal D}_b \bar{\cal D}^p {\cal E}_{ap} - \ftb{1}{2} g^{jl} \bar{\cal D}_b {\cal E}_{aj} \Big( {\cal D}^p {\cal E}_{pl} + \bar{\cal D}^p {\cal E}_{lp}  \Big) \bigg] \, .
\eq
Then by summing up all the terms, one gets the following result:
\bq
\ba{lcl}
{\cal R}_{a \bar{b} }  &=&  -  \ftb{1}{2}  ( \bar{\nabla}^s \bar{\cal D}_s {\cal E}_{a b} -  \nabla ^s {\cal D}_a {\cal E}_{sb} -  \bar{\nabla}^s \bar{\cal D}_b {\cal E}_{as})  +  (\bar{\nabla}_b {\cal D}_a d   + \nabla _a \bar{\cal D}_b d) \\ [2.0ex]
&&-   \ftb{1}{4}  g^{jk} ({\cal D}_a {\cal E}_{jb} \bar{\cal D}^i {\cal E}_{ki}  +  {\cal D}_a {\cal E}_{ki} \bar{\cal D}^i {\cal E}_{jb})   + \ftb{1}{4}  (\bar{\cal D}^j {\cal E}_{ib} {\cal D}^i {\cal E}_{aj}  +  \bar{\cal D}^i {\cal E}_{ab} {\cal D}^j {\cal E}_{ji}) \\ [2.0ex]
&&- \ftb{1}{4}  g^{jk} \bar{\cal D}_b {\cal E}_{aj} {\cal D}^i {\cal E}_{ik}    + \ftb{1}{4} g^{ik} g^{jl} \bar{\cal D}_b {\cal E}_{ij} {\cal D}_a {\cal E}_{kl} \\ [2.5ex]
&=& -  {\cal K}_{a \bar{b} } \ ,
\ea
\eq
where ${\cal K}$ in the form of \eqref{kpq2} is used. Since different forms of ${\cal K}_{pq}$ are equivalent up to the constraint, ${\cal R}_{a \bar{b}}$ defined in~\cite{Hohm:2010pp} is exactly our $- {\cal K}_{a \bar{b}}$.
\subsection{Gauge transformation of ${\cal K}_{pq}$ from generalized metric formulation}
The gauge transformation \eqref{fullgt_kpq} of ${\cal K}_{pq}$ was derived in section \ref{gaugetr section} using the gauge invariance of the double field theory action \eqref{action}. It was also explicitly checked from perturbative and non-perturbative calculations. Since we showed that ${\cal R}_{a\bar{b}} = -{\cal K}_{a \bar{b}}$  after gauge fixing, the gauge transformation of ${\cal K}_{pq}$ can be derived from the gauge transformation property of ${\cal R}_{a \bar{b}}$ in the generalized metric formulation. \\
\\
Here the gauge transformation properties of ${\cal R}_{a \bar{b}}$ will be investigated using the property that  ${\cal R}_{MN}$ is a generalized tensor. From the definition of ${\cal R}_{a \bar{b}} \equiv {e^M}_{a} {\cal R}_{MN} {e^N}_{\bar{b}}$, the gauge transformation of ${\cal R}_{a \bar{b}}$ can be written as
\bq \label{genmet_gauger}
\delta {\cal R}_{a \bar{b}} = \delta \Big( {e^M}_{a} \Big){\cal R}_{MN} {e^N}_{\bar{b}}  + {e^M}_{a} \delta \Big( {\cal R}_{MN} \Big) {e^N}_{\bar{b}} +  {e^M}_{a} {\cal R}_{MN} \delta \Big( {e^N}_{\bar{b}} \Big) \, .
\eq
Since ${\cal R}_{MN}$ is a generalized tensor i.e. $\delta {\cal R}_{MN} = \hat{\cal L}_{\xi} {\cal R}_{MN}$, the second term of \eqref{genmet_gauger} is determined. Other terms can be obtained from the gauge transformation of the frame fields. One can write the gauge and $GL (D, \mathbb{R}) \times GL (D, \mathbb{R})$ transformations of the frame fields ${e^M}_{a}$ and $ {e^N}_{\bar{b}}$ as follows:
\be
\delta {e^M}_{a} &=& \hat{\cal L}_{\xi} {e^M}_{a} + {e^M}_{c} {\Sigma^{c}}_a  \, , \\
\delta {e^N}_{\bar{b}} &=&  \hat{\cal L}_{\xi} {e^N}_{\bar{b}} + {e^N}_{\bar{c}} {\Sigma^{\bar{c}}}_{\bar{b}} \ \ .
\ee
We need $\delta  {e^i}_{a} =  \delta  {e^j}_{\bar{b}} =0$ in order to preserve the gauge condition. Then the $GL (D, \mathbb{R})$ parameters become
\be
 {\Sigma^{c}}_a &=& {\cal D}_a \xi^c - \tilde{\partial}^c \tilde{\xi}_a + \tilde{\partial}^c \xi^k {\cal E}_{ak} \, , \\
 {\Sigma^{\bar{c}}}_{\bar{b}} &=&  \bar{\cal D}_{\bar{b}} \xi^{\bar{c}} - \tilde{\partial}^{\bar{c}} \tilde{\xi}_{\bar{b}} - \tilde{\partial}^{\bar{c}} \xi^k {\cal E}_{k \bar{b}} \ \, .
\ee
Then with a little manipulation, the gauge transformation of ${\cal R}_{a \bar{b}}$ can be calculated as follows:
\bq
\ba{lcl}
\delta {\cal R}_{a \bar{b}} &=& \Big( \hat{\cal L}_{\xi}   {e^M}_{a} \Big){\cal R}_{MN} {e^N}_{\bar{b}} + {e^M}_{a} \Big( \hat{\cal L}_{\xi} {\cal R}_{MN} \Big) {e^N}_{\bar{b}} + {e^M}_{a}{\cal R}_{MN}   \hat{\cal L}_{\xi}  {e^N}_{\bar{b}}   \\ [2.0ex]
  && +  {\Sigma^{c}}_a {\cal R}_{c \bar{b}} +  {\Sigma^{\bar{c}}}_{\bar{b}} {\cal R}_{a \bar{c}} \\ [1.0ex]
&=& \xi^M \partial_M {\cal R}_{a \bar{b}} + \bigg( {\cal D}_a \xi^c - \tilde{\partial}^c \tilde{\xi}_a + \tilde{\partial}^c \xi^k {\cal E}_{ak} \bigg) {\cal R}_{c \bar{b}} + \bigg(  \bar{\cal D}_{\bar{b}} \xi^{\bar{c}} - \tilde{\partial}^{\bar{c}} \tilde{\xi}_{\bar{b}} - \tilde{\partial}^{\bar{c}} \xi^k {\cal E}_{k \bar{b}}   \bigg) {\cal R}_{a \bar{c}} \\[2.0ex]
&=&  {\cal L}_{\xi}  {\cal R}_{a \bar{b}} + {\cal L}_{\tilde{\xi}}  {\cal R}_{a \bar{b}} + {\cal E}_{ak}T^{kc}  {\cal R}_{c \bar{b}} + {\cal E}_{k \bar{b}}T^{ck}  {\cal R}_{a \bar{c}} \\ [2.0ex]
&=&  {\cal L}_{\xi}  {\cal R}_{a \bar{b}} + {\cal L}_{\tilde{\xi}}  {\cal R}_{a \bar{b}} + {\cal E}_{ak}   (\tilde{\partial}^l \xi^k - \tilde{\partial}^k \xi^l) {\cal R}_{l \bar{b}} + {\cal E}_{l \bar{b}}  (\tilde{\partial}^l \xi^k - \tilde{\partial}^k \xi^l)  {\cal R}_{a \bar{k}} \, ,
\ea
\eq
which is the same as the gauge transformation of ${\cal K}_{a \bar{b}}$ since ${\cal R}_{a \bar{b}} = - {\cal K}_{a \bar{b}}$. This result from the generalized metric formulation is another way to derive the gauge transformation of ${\cal K}_{pq}$.
\section{Conclusions}
In this paper we have thoroughly investigated the invariances and equations of motion in background independent double field theory~\cite{Hohm:2010jy}. The Ricci-like tensor ${\cal K}_{pq}$ has been derived from the ${\cal E}$-variation of the action \eqref{action} and is an $O(D,D)$ covariant tensor. The gauge transformation of ${\cal K}_{pq}$ is given by \eqref{fullgt_kpq} and can be also written using $O(D,D)$ covariant derivatives and parameters \eqref{bi_gt}. However, $\delta {\cal K}_{pq}$ is not an $O(D,D)$ covariant tensor. From the theorem we have constructed, it is a general result that the gauge transformation of an $O(D,D)$ tensor (not a scalar) is not an $O(D,D)$ tensor.\\
\\
Our results were compared with those in General Relativity. Contracted Bianchi identities in double field theory, derived from the gauge invariance of the action, have a structure similar to that of General Relativity. In the limit of no $\tilde{x}$ dependence, they exactly reduce to the Bianchi identity of General Relativity. Similarly, the equations of motion ${\cal R}=0$ and ${\cal K}_{pq}=0$ reduce to those in a conventional GR setup in this limit. \\
\\
We examined the gauge transformation property of ${\cal R}_{MN}$ and found that it is a generalized tensor using the gauge invariance of the action and the gauge transformation of ${\cal K}_{MN}$. Then we investigated the equivalence between ${\cal R}_{a \bar{b}}$ and ${\cal K}_{a \bar{b}}$. Using these properties, we could derive the gauge transformation of ${\cal K}_{pq}$ in the generalized metric formulation. 
\subsection*{Acknowledgments}
I am happy to acknowledge very helpful discussions with Barton Zwiebach and Olaf Hohm. I am truly thankful to Barton Zwiebach for his instructions and advices for this work. This work is supported by U.S. Department of Energy (D.O.E.) under cooperative research arrangement DE-FG02-05ER41360. The work of SK is supported in part by Samsung Scholarship.
\appendix
\section{Explicit Calculation of ${\cal K}^{pq}$} \label{appA}
We will vary the action \eqref{symaction} with respect to $\cal E$, starting from the last term of the action and proceed in the reverse order. Note that all equalities are up to integration by parts here and the constraint are used in a few places to simplify terms.\\
First by varying the last term of the action \eqref{symaction}, one can obtain
\bq \label{easiest}
\ba{lcl}
\delta_{\cal E} (2 e^{-2d}  {\cal D}^i d \ {\cal D}_i d ) &=& 2 e^{-2d} \bigg[ \delta_{\cal E} (g^{ij} ){\cal D}_i d {\cal D}_j d + \delta {\cal E}_{ik} g^{ij} {\cal D}_j d  ({\cal D}^k d - \bar{\cal D}^k d)  \bigg] \\ [2.0ex]
&=& 2 e^{-2d} \bigg[ \delta {\cal E}_{pq} {\cal D}^p d {\cal D}^q d +  \delta {\cal E}_{pq} {\cal D}^p ({\cal D}^q d - \bar{\cal D}^q d)  d \bigg]  \\ [3.0ex]
&=& \delta {\cal E}_{pq} e^{-2d} \big[ -2 {\cal D}^p d \ \bar{\cal D}^q d \ \big] = \delta_{\cal E} (2 e^{-2d}  \bar{\cal D}^i d \ \bar{\cal D}_i d ) \, .
\ea
\eq
All other terms are varied in the same way, hence we do not display detailed calculation for other terms. By varying the second and third term in the second line of \eqref{symaction}, one finds
\bq
\ba {lcl} \label{nexteasy}
\delta_{\cal E} \bigg[ e^{-2d}( {\cal D}^i d \  \bar{\cal D}^j {\cal E}_{ij} + \bar{\cal D}^i d \ {\cal D}^j {\cal E}_{ji} ) \bigg] &=& \delta {\cal E}_{pq}  e^{-2d} \bigg[ -(\bar{\nabla}^q {\cal D}^p d + \nabla^p \bar{\cal D}^q d)- g^{qj} {\cal D}^i d \ {\cal D}^p {\cal E}_{ij}  \\ 
&&- \, g^{pi} \bar{\cal D}^j d \ \bar{\cal D}^q {\cal E}_{ij} + g^{pk} g^{ql}  {\cal D}^s d \ {\cal D}_s {\cal E}_{kl}  + 4 {\cal D}^p d \ \bar{\cal D}^q d \ \bigg] \, ,
\ea
\eq
To evaluate the rest terms now define
\bq
\ba{lcl}
A \equiv e^{-2d}  g^{ik} g^{jl} {\cal D}^p {\cal E}_{kl} {\cal D}_p {\cal E}_{ij}  &,& \bar{A} \equiv e^{-2d} g^{ik} g^{jl} \bar{\cal D}^p {\cal E}_{kl} \bar{\cal D}_p {\cal E}_{ij} \, , \\  [1.5ex]
B \equiv e^{-2d} g^{kl} {\cal D}^j {\cal E}_{ik} {\cal D}^i {\cal E}_{jl}  &,& C \equiv e^{-2d} g^{kl} \bar{{\cal D}}^j {\cal E}_{ki} \bar{\cal D}^i {\cal E}_{lj} \, .
\ea
\eq
The variation of the above equations gives the following results:
\bq
\ba{c}
 \delta_{\cal E} A =   \delta {\cal E}_{pq}  e^{-2d} \bigg[  -2 g^{ip} g^{jq} \nabla^s {\cal D}_s {\cal E}_{ij} - g^{pl} g^{jk} \bar{\cal D}^q {\cal E}_{ik} {\cal D}^i {\cal E}_{lj}  + g^{pk} g^{ql} \bar{\cal D}^j {\cal E }_{il} {\cal D}^i {\cal E}_{kj} \\ 
  - g^{ik} g^{jl} {\cal D}^p {\cal E}_{kl} \bar{\cal D}^q {\cal E}_{ij}  +  g^{pk} g^{ql} \bar{\cal D}^{i} {\cal E}_{ji} {\cal D}^j {\cal E}_{kl} + 4 g^{pk} g^{ql} {\cal D}^s d \ {\cal D}_s {\cal E}_{kl}  \ \bigg] \, ,
\ea
\eq
\bq
\ba{c}
\delta_{\cal E} \bar{A} = \delta {\cal E}_{pq}  e^{-2d} \bigg[   -2 g^{ip} g^{jq} \bar{\nabla}^s \bar{\cal D}_s {\cal E}_{ij} - g^{ql} g^{jk} {\cal D}^p {\cal E}_{ki} \bar{\cal D}^{i} {\cal E}_{jl} + g^{pk} g^{ql} \bar{\cal D}^j {\cal E }_{il} {\cal D}^i {\cal E}_{kj} \\ - g^{ik} g^{jl} {\cal D}^p {\cal E}_{kl} \bar{\cal D}^q {\cal E}_{ij}  + g^{pk} g^{ql} \bar{\cal D}^i {\cal E}_{kl} {\cal D}^j {\cal E}_{ji} + 4 g^{pk} g^{ql} \bar{\cal D}^s d \ \bar{\cal D}_s {\cal E}_{kl}  \ \bigg] \, ,
\ea
\eq
\bq
\ba{c}
 \delta_{\cal E} B =   \delta {\cal E}_{pq}  e^{-2d}  \bigg[  -2 g^{ql} \nabla^j {\cal D}^p {\cal E}_{jl} - g^{ik} g^{jl} {\cal D}^p {\cal E}_{kl} \bar{\cal D}^q {\cal E}_{ij}  - g^{kl} g^{jp} \bar{\cal D}^q {\cal E }_{ik} {\cal D}^i {\cal E}_{jl} \\ + g^{kl} g^{jq} {\cal D}^p {\cal E}_{ki} \bar{\cal D}^i {\cal E}_{lj} +  g^{ql} g^{jk} \bar{\cal D}^{i} {\cal E}_{ki} {\cal D}^p {\cal E}_{jl} + 4 g^{qj} {\cal D}^i d \ {\cal D}^p {\cal E}_{ij} \ \bigg] \, ,
\ea
\eq
\bq
\ba{c}
 \delta_{\cal E} C =   \delta {\cal E}_{pq}  e^{-2d} \bigg[  -2 g^{pl} \bar{\nabla}^j \bar{\cal D}^q {\cal E}_{lj} - g^{ik} g^{jl} {\cal D}^p {\cal E}_{kl} \bar{\cal D}^q {\cal E}_{ij}  + g^{kl} g^{jp} \bar{\cal D}^q {\cal E }_{ik} {\cal D}^i {\cal E}_{jl} \\ - g^{kl} g^{jq} {\cal D}^p {\cal E}_{ki} \bar{\cal D}^i {\cal E}_{lj} +  g^{pl} g^{jk} {\cal D}^{i} {\cal E}_{ik} \bar{\cal D}^q {\cal E}_{lj} +4 g^{pi} \bar{\cal D}^j d \ \bar{\cal D}^q {\cal E}_{ij} \ \bigg] \, .
\ea
\eq
Then the variation of the first four terms of \eqref{symaction} can be written as
\bq \label{firstline}
\ftb{1}{4} \bigg[ -\ftb{1}{2} \Big(  \delta_{\cal E} A + \delta_{\cal E} \bar{A}  \Big) + \delta_{\cal E} B + \delta_{\cal E} C \bigg] \, .
\eq       
By combining \eqref{firstline}, \eqref{easiest}, and \eqref{nexteasy}, one obtains \eqref{kpq3}. \\
\\
In section \ref{checkrandk} the verification of the equivalence ${\cal R}_{a \bar{b}} = - {\cal K}_{a \bar{b}}$ used the following definition:
\bq \label{kpq2}
\ba{lcl}
{\cal K}^{pq} &=&  \ftb{1}{2} g^{ip} g^{jq} ( \bar{\nabla}^s \bar{\cal D}_s {\cal E}_{ij} -  \nabla ^s {\cal D}_i {\cal E}_{sj} -  \bar{\nabla}^s \bar{\cal D}_j {\cal E}_{is})  - g^{pi} g^{qj} (\bar{\nabla}_j {\cal D}_i d   + \nabla _i \bar{\cal D}_j d) \\ [2.0ex]
&&  +  \  \ftb{1}{4} g^{ql} g^{jk} ({\cal D}^p {\cal E}_{jl} \bar{\cal D}^i {\cal E}_{ki}  +  {\cal D}^p {\cal E}_{ki} \bar{\cal D}^i {\cal E}_{jl})   - \ftb{1}{4} g^{pk} g^{ql} (\bar{\cal D}^j {\cal E}_{il} {\cal D}^i {\cal E}_{kj}  +  \bar{\cal D}^i {\cal E}_{kl} {\cal D}^j {\cal E}_{ji}) \\[2.0ex]
&&  + \ \ftb{1}{4} g^{pl} g^{jk} \bar{\cal D}^q {\cal E}_{lj} {\cal D}^i {\cal E}_{ik}    - \ftb{1}{4} g^{ik} g^{jl} \bar{\cal D}^q {\cal E}_{ij} {\cal D}^p {\cal E}_{kl} \, ,
\ea
\eq
which is the $\cal E$-variation of the following action:
\bq
\ba {c}
S = \displaystyle\int dx d \tilde{x}  \ e^{-2d} \Bigg[ - \ftb{1}{4} g^{ik} g^{jl} \bar{\cal D}^p {\cal E}_{kl} \bar{\cal D}_p {\cal E}_{ij} + \ftb{1}{4} g^{kl} ({\cal D}^j {\cal E}_{ik} {\cal D}^i {\cal E}_{jl} + \bar{{\cal D}}^j {\cal E}_{ki} \bar{\cal D}^i {\cal E}_{lj}) \\ +  ({\cal D}^i d \  \bar{\cal D}^j {\cal E}_{ij} + \bar{\cal D}^i d \ {\cal D}^j {\cal E}_{ji} ) + 4 {\cal D}^i d \ {\cal D}_i d \ \Bigg] \, .
\ea
\eq
This action is equivalent to the actions \eqref{action} and \eqref{symaction} using the constraint, thus ${\cal K}^{pq}$ defined in \eqref{kpq2} is equivalent to that defined in \eqref{kpq3}.

\section{Proof of general $O(D,D)$ structure of $\cal E$-variation} \label{genoddthm1}
Before starting the proof, note that the ${\cal E}$-variation of ${\cal D}d$, $\bar{\cal D}d$, ${\cal D}{\cal E}$, $\bar{\cal D}{\cal E}$, and $g^{ij}$ satisfy the claimed structure in the theorem; we have checked this explicitly. The proof consists of three steps:\\
\\ 
\textbf{Step I.} First we will consider an $O(D,D)$ tensor $T$ only with lower indices (i.e. no contractions of upper and lower indices      or raised index with $g^{ij}$). Then we will show that the ${\cal E}$-variation of $T$ has the following $O(D,D)$ structure:
\bq \label{tensorprop} 
\delta_{\cal E} T_{i_1 i_2 \cdots i_n} = \bigg(O(D, D) \text{ terms} \bigg) +\displaystyle\sum_{\substack{\text{all unbarred} \\ \text{indices $k$}}} \ftb{1}{2} \delta {\cal E}_{kq}g^{ql} T_{\{ k \rightarrow l \}} +  \displaystyle\sum_{\substack{\text{all barred} \\ \text{indices $k$}}} \ftb{1}{2} \delta {\cal E}_{pk} g^{pl} T_{\{ k \rightarrow l\}} \, ,
\eq 
where $T_{ \{ k \rightarrow l \} }$ is the tensor $T$ with index $l$ substituted for index $k$. \\
\\
Before starting the proof, note that one can always use a reordering operation for the indices of a tensor $T$ to define the following  tensor:
\bq \label{reord}
\tilde{T}_{j_1 j_2 \cdots j_m \bar{j}_1 \bar{j}_2 \cdots \bar{j}_l} \equiv T_{i_1 i_2 \cdots i_n} \, ,
\eq
where indices $i_k$ can be either unbarred or barred and $m+l =n$. This reordering operation sends all unbarred indices of $T$ to the front. Then for this tensor $\tilde{T}$, the formula \eqref{tensorprop} becomes
\bq \label{ord_tensor_prop}
\ba{lcl}
\delta_{\cal E} \tilde{T}_{j_1 \cdots j_m \bar{j}_1  \cdots \bar{j}_l} = \displaystyle\sum_{j_k}  \ftb{1}{2} \delta {\cal E}_{j_k q}g^{ql} \tilde{T}_{j_1 \cdots j_{k-1} \ l \ j_{k+1} \cdots \bar{j}_l}  
+ \displaystyle\sum_{\bar{j}_k} \ftb{1}{2} \delta {\cal E}_{p \bar{j}_k } g^{pl} \tilde{T}_{j_1 \cdots  \bar{j}_{k-1} \ l \ \bar{j}_{k+1} \cdots \bar{j}_l} + \cdots \, ,
\ea
\eq 
where dots indicate $O(D,D)$ covariant terms. If the tensor $T_{i_1 i_2 \cdots i_n}$ only consists of ${\cal D}d$, $\bar{\cal D}d$, ${\cal D}{\cal E}$, and $\bar{\cal D}{\cal E}$, then the tensor automatically satisfy the formula \eqref{tensorprop}. Now the only other possibility of forming an $O(D, D)$ tensor is to include covariant derivatives $\nabla$ and $\bar{\nabla}$. We will prove that this type of tensor also obeys the formula \eqref{tensorprop} by induction.\\
\\
Suppose that a tensor $T$ satisfies the formula \eqref{tensorprop}. After reordering, consider the following variation:
\bq \label{varnabla}
\delta_{\cal E} \nabla_i \tilde{T}_{j_1 j_2 \cdots j_m \bar{j}_1 \bar{j}_2 \cdots \bar{j}_l} =\Big( \delta_{\cal E} \nabla_i \Big) \tilde{T}_{j_1 j_2 \cdots j_m \bar{j}_1 \bar{j}_2 \cdots \bar{j}_l}
+ \nabla_i \Big( \delta_{\cal E}  \tilde{T}_{j_1 j_2 \cdots j_m \bar{j}_1 \bar{j}_2 \cdots \bar{j}_l} \Big) \, ,
\eq
where $( \delta_{\cal E} {\nabla_i} )$ means the $\cal E$-variation of the calligraphic derivative and Christoffel-like symbols in the $O(D,D)$ covariant derivative $\nabla$. The variation \eqref{varnabla} sums up to
\bq
\ba {lcl}
\delta_{\cal E} \nabla_i \tilde{T}_{j_1 j_2 \cdots j_m \bar{j}_1 \bar{j}_2 \cdots \bar{j}_l}  &=& \bigg(O(D, D) \text{ terms} \bigg) + \Bigg[ \ftb{1}{2} \delta {\cal E}_{iq} g^{ql} \nabla_l \tilde{T}  + \displaystyle\sum_{j_k}  \ftb{1}{2} \delta {\cal E}_{j_k q}g^{ql} \nabla_i   \tilde{T}_{\{ j_k \rightarrow l \}} \Bigg] \\
&&+\displaystyle\sum_{\bar{j}_k}  \ftb{1}{2}  \delta {\cal E}_{ p \bar{j}_k }g^{pl} \nabla_i \tilde{T}_{\{ \bar{j}_k \rightarrow l \}}  \, ,
\ea
\eq
which satisfies the formula \eqref{ord_tensor_prop}. Thus $\delta_{\cal E} \nabla_i T$ satisfies the formula \eqref{tensorprop} by reordering the indices. \\
\\
For $\bar{\nabla}$, we can apply the same procedure and the results satisfy the formula \eqref{tensorprop}. \\ 
Since $T$ only consisting of ${\cal D}d$, $\bar{\cal D}d$, ${\cal D}{\cal E}$, and $\bar{\cal D}{\cal E}$ satisfies the formula \eqref{tensorprop}, $\nabla_i T$ also satisfies  \eqref{tensorprop} for such $T$. Then the ${\cal E}$-variation of any $O(D,D)$ tensor $T$ only with lower indices satisfies \eqref{tensorprop} by induction. Note that in this step we did not include any contraction between indices.\\
\\ 
\textbf{Step II.} In this step we will show that any $O(D, D)$ contraction (i.e. contraction with $g^{ij}$ and $g^{\bar{i}\bar{j}}$) of the lower indices does not change the claimed structure. Here we will assume that a tensor $T$ with only lower indices satisfy the formula \eqref{tensorprop}. Consider some $O(D, D)$ contraction between two indices of $T$. Using reordering operation, we can rearrange these two indices to be the first two indices. Then the variation of this contraction gives (for unbarred indices and barred indices respectively)
\be
\delta_{\cal E} g^{ik} \tilde{T}_{ik j_1 \cdots} = \bigg(O(D, D) \text{ terms} \bigg) +\displaystyle\sum_{j_k}  \ftb{1}{2} \delta {\cal E}_{j_k q}g^{ql}  \Big(g^{ik} \tilde{T}_{\{ j_k \rightarrow l \}} \Big)  + \displaystyle\sum_{\bar{j}_k} \ftb{1}{2} \delta {\cal E}_{p \bar{j}_k } g^{pl} \Big( {g^{ik}} \tilde{T}_{\{ \bar{j}_k \rightarrow l \}} \Big) \, , \\
\delta_{\cal E} g^{ik} \tilde{T}_{\bar{i} \bar{k} j_1 \cdots }  = \bigg(O(D, D) \text{ terms} \bigg) +\displaystyle\sum_{j_k}  \ftb{1}{2} \delta {\cal E}_{j_k q}g^{ql} \Big( g^{ik} \tilde{T}_{\{ j_k \rightarrow l \}} \Big)   + \displaystyle\sum_{\bar{j}_k} \ftb{1}{2} \delta {\cal E}_{p \bar{j}_k } g^{pl} \Big( {g^{ik}} \tilde{T}_{\{ \bar{j}_k \rightarrow l \}} \Big) \, ,
\ee
which satisfy the formula \eqref{tensorprop}. Note that since $\nabla g^{ij} = \bar{\nabla} g^{ij} = \nabla g^{\bar{i} \bar{j}} = \bar{\nabla} g^{\bar{i} \bar{j}} = 0$, any contraction inside covariant derivatives does not affect this formula.  \\
\\ 
\textbf{Step III.} Here consider a general $O(D, D)$ tensor $T_{i_1 i_2 \cdots i_l}^{i_1' i_2 ' \cdots i_{l'}'}$. Using a reordering operation, one can define
\bq
{\tilde{T}^{k_1 k_2 \cdots k_{m'} \bar{k}_1 \bar{k}_2 \cdots \bar{k}_{n'}}}_{j_1 j_2 \cdots j_{m} \bar{j}_1 \bar{j}_2 \cdots \bar{j}_n} \equiv T_{i_1 i_2 \cdots i_l}^{i_1' i_2 ' \cdots i_{l'}'} \, ,
\eq
where $i_k$, $i_k '$ can be either unbarred and barred and $m+n = l$, $m' + n' = l'$. This reordering operation is defined in a similar way to \eqref{reord}. Then we can write this tensor as follows:
\bq
{\tilde{T}^{k_1 k_2 \cdots k_{m'} \bar{k}_1 \bar{k}_2 \cdots \bar{k}_{n'}}}_{j_1 j_2 \cdots j_{m} \bar{j}_1 \bar{j}_2 \cdots \bar{j}_n} = g^{k_1 s_1} g^{k_2 s_2} \cdots g^{k_{m'} s_{m'}} g^{\bar{k}_1 \bar{s}_1} \cdots g^{\bar{k}_{n'} \bar{s}_{n'}} \tilde{T}^{\ast}_{j_1 \cdots j_m s_1 \cdots s_{m'} \bar{j}_1 \cdots \bar{j}_n \bar{s}_1 \cdots \bar{s}_{n'}} \, .
\eq
The variation of this tensor is given by
\bq
\ba{lcl}
\delta_{\cal E} {\tilde{T}^{k_1 k_2 \cdots k_{m'} \bar{k}_1 \bar{k}_2 \cdots \bar{k}_{n'}}}_{j_1 j_2 \cdots j_{m} \bar{j}_1 \bar{j}_2 \cdots \bar{j}_n} &=& \bigg(O(D, D) \text{ terms} \bigg)  +\displaystyle\sum_{j_k}  \ftb{1}{2} \delta {\cal E}_{j_k q}g^{ql} \tilde{T}_{\{ j_k \rightarrow l \}}   
+ \displaystyle\sum_{\bar{j}_k} \ftb{1}{2} \delta {\cal E}_{p \bar{j}_k } g^{pl} \tilde{T}_{\{ \bar{j}_k \rightarrow l \}} \\
&&- \  \displaystyle\sum_{k_i}  \ftb{1}{2} \delta {\cal E}_{pq}  g^{k_i q} \tilde{T}^{ \{ k_i \rightarrow p \} }   - \displaystyle\sum_{\bar{k}_i} \ftb{1}{2} \delta {\cal E}_{pq}   g^{\bar{k}_i p} \tilde{T}^{\{ \bar{k}_i \rightarrow q \}} \, .
\ea
\eq
By recovering the order of indices using reordering operation, one can easily see that this gives the formula \eqref{tensor_realprop}. 
\section{Equation of motions and Bianchi identity in a conventional GR setup} \label{usualgr}
In a conventional GR setup, the action is written in terms of metric, $H$-field, and dilaton as
\bq
S = \int \sqrt{-g} e^{-2 \phi} \bigg[ R + 4 (\partial \phi)^2 - \ftb{1}{12} H^2 \bigg] \, .
\eq
The variation of the action with respect to $g^{ij}$, $b_{ij}$, and $\phi$ are
\bq
\ba{lcl}
\delta_{g} S &=& \displaystyle\int \sqrt{-g} e^{-2 \phi} \delta g^{ij} \bigg[ R_{ij} - \ftb{1}{2} g_{ij} R +2 \Big( -g_{ij} \nabla^2 \phi + \nabla_{i} \nabla_{j} \phi + g_{i j} (\partial \phi)^2 \Big) \\ [2.0ex]
&& - \ftb{1}{4} \Big( {H_i}^{p q} H_{j p q} - \ftb{1}{6} g_{i j} H^2 \Big) \bigg] , \\ [2.0ex] 
\delta_{b} S &=& \displaystyle\int \sqrt{-g} e^{-2 \phi} \delta b_{ij} \bigg[ \ftb{1}{2} \nabla_p H^{ijp} - H^{ijp} \nabla_p \phi \ \bigg] , \\ [2.0ex]
 \delta_{\phi} S &=& -2 \displaystyle\int \sqrt{-g} e^{-2 \phi} \delta \phi \bigg[ R+4 \Big( \nabla^2 \phi - (\partial \phi)^2 \Big) - \ftb{1}{12} H^2  \bigg] . \\
\ea
\eq
 The gauge transformations of $g^{ij}$, $b_{ij}$ and $\phi$ are as follows, respectively:
 \bq \label{gaugetrfgr}
\delta g^{ij} = - \nabla^i \xi^j - \nabla^j \xi^i \ , \quad 
\delta b_{ij} = H_{ijk} \xi^k + \partial_i \tilde{\xi}_j  - \partial_j \tilde{\xi}_i \ , \quad
\delta \phi = \xi^i \partial_i \phi \, .
\eq
Then using the integration by parts, the total variation of the action is
\bq
\ba{lcl}
\delta S &=& -2 \displaystyle\int \sqrt{-g}  e^{-2 \phi} \Bigg\{ \xi^j \bigg[ \Big( \nabla^i R_{ij} - \ftb{1}{2} \nabla_j R \Big) + \ftb{1}{2} \Big( \ftb{1}{2} {H_i}^{p q} \nabla^i H_{j p q} - \ftb{1}{12} \nabla_j H^2 \Big) \bigg] \\ && + \tilde{\xi}_j \bigg[ - \ftb{1}{2} \nabla_p \nabla_q H^{pqj} \bigg] \Bigg\} .
\ea
\eq
Using \eqref{JacobiRiemann} and \eqref{JacobiH}, this total variation gives the following Bianchi identity:
\bq
\delta S = \displaystyle\int \sqrt{-g} e^{-2 \phi} \xi^j \bigg[ \nabla^i R_{ij} - \ftb{1}{2} \nabla_j R \bigg]  = 0 \quad  \Longleftrightarrow \quad \nabla^i R_{ij} - \ftb{1}{2} \nabla_j R=0 \, .
\eq
Note that the introduction of $H$-field and dilaton does not modify the Bianchi identity.


\begin{thebibliography}{99}

\baselineskip 15pt

\bibitem{Giveon:1994fu}
  A.~Giveon, M.~Porrati and E.~Rabinovici,
  ``Target space duality in string theory,''
  Phys.\ Rept.\  {\bf 244}, 77 (1994)
  [arXiv:hep-th/9401139].

\bibitem{Hull:2009mi}
  C.~Hull and B.~Zwiebach,
  ``Double Field Theory,''
  JHEP {\bf 0909} (2009) 099
  [arXiv:0904.4664 [hep-th]].

\bibitem{Hull:2009zb}
  C.~Hull and B.~Zwiebach,
  ``The gauge algebra of double field theory and Courant brackets,''
  JHEP {\bf 0909} (2009) 090
  [arXiv:0908.1792 [hep-th]].

\bibitem{Hohm:2010jy}
  O.~Hohm, C.~Hull and B.~Zwiebach,
  ``Background independent action for double field theory,''
  JHEP {\bf 1007}, 016 (2010)
  [arXiv:1003.5027 [hep-th]].


\bibitem{Hohm:2010pp}
  O.~Hohm, C.~Hull and B.~Zwiebach,
  ``Generalized metric formulation of double field theory,''
  JHEP {\bf 1008}, 008 (2010)
  [arXiv:1006.4823 [hep-th]].

\bibitem{Kugo:1992md}
  T.~Kugo and B.~Zwiebach,
  ``Target space duality as a symmetry of string field theory,''
  Prog.\ Theor.\ Phys.\  {\bf 87}, 801 (1992)
  [arXiv:hep-th/9201040].
  
\bibitem{Zwiebach:1992ie}
  B.~Zwiebach,
  ``Closed string field theory: Quantum action and the B-V master equation,''
  Nucl.\ Phys.\  B {\bf 390}, 33 (1993)
  [arXiv:hep-th/9206084].
  
  
  \bibitem{Tseytlin:1990nb}
A.~A.~Tseytlin,
``Duality Symmetric Formulation Of String World Sheet Dynamics,''
Phys.\ Lett.\ B {\bf 242}, 163 (1990);
``Duality Symmetric Closed String Theory And Interacting Chiral Scalars,''
Nucl.\ Phys.\ B {\bf 350}, 395 (1991).




\bibitem{Siegel:1993th}
  W.~Siegel,
  ``Superspace duality in low-energy superstrings,''
  Phys.\ Rev.\  D {\bf 48}, 2826 (1993)
  [arXiv:hep-th/9305073].


\bibitem{Siegel:1993xq}
W.~Siegel,
  ``Two vierbein formalism for string inspired axionic gravity,''
  Phys.\ Rev.\  D {\bf 47}, 5453 (1993)
  [arXiv:hep-th/9302036].
  
  \bibitem{Gualtieri}
M.~Gualtieri,
``Generalized complex geometry,"
PhD Thesis (2004).
arXiv:math/0401221v1 [math.DG]

  

\end{thebibliography}
\end{document}